\documentclass[journal,12pt,onecolumn]{IEEEtran}
\usepackage{cite}
\usepackage{amsmath}
\usepackage{amsmath}
\usepackage{amsfonts}   
\usepackage{amssymb}    
\usepackage{float}
\usepackage{subfigure}
\usepackage{float}
\usepackage{amsmath}
\usepackage{color}
\usepackage{graphicx}
\usepackage{epsfig}
\usepackage{algorithm}
\usepackage{enumerate}
\usepackage[noend]{algorithmic}
\usepackage[doublespacing]{setspace}
\newcommand{\figref}[1]{Fig.~\ref{#1}}
\begin{document}
\title{An Intercell Interference Model based on Scheduling  for Future Generation Wireless Networks (Part I and Part II)}
{
\author{\IEEEauthorblockN{Hina Tabassum$^*$, Ferkan Yilmaz$^*$, Zaher Dawy$^\star$, Mohamed-Slim
Alouini$^*$} \\
}
}
\maketitle
\begin{abstract}
This technical report is divided into two parts.
The first part of the technical report presents a  novel framework for modeling the uplink and downlink intercell interference (ICI) in a multiuser cellular network. The proposed framework assists in quantifying the impact of various fading channel models and multiuser scheduling schemes on the uplink and downlink ICI. Firstly, we derive a semi-analytical expression for the distribution of the location of the scheduled user in a given cell considering a wide range of scheduling schemes. Based on this, we derive the distribution and  moment generating function (MGF) of the  ICI considering a single interfering cell. Consequently, we determine the MGF of the cumulative ICI observed from all interfering cells and derive explicit MGF expressions for three typical fading models. Finally, we utilize the obtained expressions to evaluate important network performance metrics such as the outage probability, ergodic capacity and average fairness numerically. Monte-Carlo simulation results are provided to demonstrate the efficacy of the  derived analytical expressions\footnote{\bf The first part of the technical report is currently submitted to IEEE Transactions on Wireless Communications}.

The second part of the technical report deals with the statistical modeling of uplink inter-cell interference (ICI) considering greedy scheduling with power adaptation based on channel conditions. 
The derived model is utilized to evaluate important network performance metrics such as  ergodic capacity, average fairness and average power preservation  numerically. In parallel to the literature, we have shown that greedy scheduling with power adaptation reduces the ICI, average power consumption of users, and enhances the average fairness among users, compared to the case without power adaptation. 
\end{abstract}
\IEEEpeerreviewmaketitle

\section*{\bf Part I of the Technical Report}
\section{Part I: Introduction}

Explosive growth in the demand of high quality wireless data services compel the  network  designers to utilize spectrum more aggressively which on one side enhances the spectrum efficiency, whereas on the other side it enhances the
intercell interference (ICI) which  is  an alarming bottleneck in the telecommunication growth paradigm. 
The allocation of the same frequency bands across neighboring cells produces indeterministic ICI which is highly dependent on the statistics of the channel characteristics and on the dynamics of the multiuser scheduling decisions. 
In this context,  it is of immense importance for the system designers to accurately characterize and investigate the behavior of the ICI  which  helps in gaining more theoretical insights,  quantifying various network performance metrics and developing efficient resource allocation and interference mitigation schemes.

Orthogonal frequency division multiple access (OFDMA) has been recently adopted as the multiple access scheme for the state-of-the-art LTE and WiMAX cellular technologies. In OFDMA, a wide-band frequency-selective fading channel is decomposed into a set of orthogonal narrow-band  subcarriers. The orthogonality among the subcarriers per cell makes the intra-cell interference almost negligible. However,  with universal frequency reuse among cells (i.e., all cells use the same set of subcarriers), the ICI at each subcarrier may cause severe degradation in the network performance. In OFDMA networks, the subcarriers are allocated adaptively among users per cell based on a predefined scheduling scheme. Moreover, each subcarrier is allocated to only one user per cell at a given time instant and, thus, the number of interfering users on each subcarrier is rather limited. Therefore, the cumulative ICI on a given subcarrier may not be modeled accurately as a Gaussian random variable (RV) by simply invoking the central limit theorem.

Several recent studies considered the modeling of ICI in the downlink where the location of interferers is typically deterministic. In \cite{arxiv}, a semi-analytical approach to estimate the ICI by modifying the Burr distribution considering path loss, Rayleigh fading and log-normal shadowing was presented. A semi-analytical distribution for the signal-to-interference-noise ratio (SINR) has been derived in \cite{eurasip} under path loss and log normal shadowing for randomly located femtocell networks. In \cite{plass}, the applicability of the Gaussian and binomial distributions for modeling the downlink ICI is investigated. In \cite{yang},  the optimal threshold is derived for fractional frequency reuse (FFR) systems  assuming ICI as Gaussian RV. In \cite{TWC1}, the authors derived the distribution of the ICI under log-normal shadowing and Rayleigh fading. The distribution of ICI is shown to highly deviate from the Gaussian distribution in OFDMA networks.

In comparison to downlink, the nature of uplink intercell interference (ICI) is different in various aspects that include the following: 
(i) Due to the implicit symmetry and fixed locations of the BSs in the typical grid-based downlink network models, the number of significantly contributing interferers typically remains the same irrespective of the position of the mobile receiver. Also, it has been shown in \cite{plass} that the strongest interference is generated by  two closest interfering BSs  irrespective of the mobile receiver location. However, the number of significantly contributing interferers in the uplink cannot be quantified at a given instant due to the highly varying locations of the interfering mobile transmitters;
(ii) Conditioned on the location of the desired mobile receiver within a cell, the exact  distance of the interfering BSs can be calculated in the typical grid-based downlink network models. However, knowing  the location of the BS receiver in the uplink  does not help in determining the exact location of the interfering mobile users;
(iii) In the uplink, cell edge and cell center mobile users are subject to the same amount of interference on a given subcarrier, which is the interference received at the BS.  Whereas the same is not true for the downlink  in which cell edge  users  experience higher interference coming from the nearby  BSs \cite{TWC,viering}.

From the  system design perspective and  performance analysis, both uplink and downlink are   equally important for the network designers. However, based on the above arguments,  the nature of the uplink ICI on a given subcarrier is different from the downlink ICI due to the inherit differences in the nature of the interferers. Therefore, uplink and downlink require different modeling approaches to statistically capture their interference dynamics.
In this context, the proposed analytical approach to model uplink ICI on a given subcarrier is novel  as it captures the impact of various state-of-the-art  scheduling schemes and is generalized for different composite fading models. The approach provides the MGF of the ICI which can be utilized to calibrate various system performance metrics.

Some worth mentioning  research works for the uplink appear in \cite{TWC, IEEEletter,rup,ICC}.
In \cite{TWC}, the authors developed an analytical model for subcarrier collisions as a function of the cell load and frequency reuse pattern. They derived an expression for the SINR in the uplink and downlink, ignoring the effect of shadowing and fading. In \cite{IEEEletter}, the authors developed an analytical expression for the subcarrier collision probability considering non-coordinated schedulers. In \cite{rup}, the authors modeled uplink ICI in an OFDMA network as a function of the reuse partitioning radius and traffic load assuming arbitrary scheduling. In \cite{ICC}, the authors presented a semi-analytical method to approximate the distribution of the uplink ICI through numerical simulations without considering the impact of scheduling schemes.

In this paper, we propose a novel theoretical framework to derive the statistics of  the uplink and downlink ICI  on a given subcarrier as a function of both the channel statistics (i.e., path loss, shadowing and fading) and multiuser scheduling decisions. The framework is generic in the sense that the derivations hold for generalized fading channels and various scheduling algorithms.   We start by deriving the distribution of the location of the scheduled user in a given cell. We then derive the distribution and moment generating function (MGF) of the ICI considering a single interfering cell. Next, we derive the MGF expression for the cumulative ICI experienced from all interfering cells over generalized fading channels, and present explicit expressions for three practical fading models. Finally, we demonstrate the importance of the derived expressions by utilizing them to evaluate important network performance metrics.

\noindent\textbf{Notation}:  $\mathrm{Exp}(\lambda)$ represents an exponential distribution with parameter $\lambda$, $\mathrm{Gamma}(m_s,m_c)$ represents a Gamma distribution with shape parameter $m_s$ and scale parameter $m_c$. $\mathcal{K}_G (m_c,m_s,\Omega)$ represents the Generalized-$\mathcal{K}$ distribution with  fading parameter $m_c$, shadowing parameter $m_s$ and average power $\Omega$. $\Gamma(.)$ represents the Gamma function. $p(A)$ denotes the probability of event A. $f(.)$ and $F(.)$ denotes the probability distribution function (PDF) and cumulative distribution function (CDF), respectively. $[a,b]$ denotes a discrete set of elements which ranges from $a$ to $b$. Finally, $\mathbb{E}[.]$ denotes the expectation operator.

\section{Part I: System Model and Proposed Framework}
\subsection{Description of the System Model}
We consider a given cell surrounded by $L$ interfering neighboring cells. For analytical convenience, the cells are assumed to be circular with radius $R$. Each cell $l$ is assumed to have $U$ uniformly distributed users. 
The frequency reuse factor is assumed to be unity with each subcarrier reused in all cells.
The bandwidth of a subcarrier is assumed to be less than the channel coherence bandwidth, thus, each subcarrier experiences flat fading. Time is divided into time slots of length smaller than the channel coherence time and, thus, the channel variation within a given time slot is negligible.

Generally, the scheduling strategies can be broadly categorized into two classes; (i) rate maximization (i.e., rate adaptation) while transmitting with  constant/maximum power; (ii) power minimization (i.e., power adaptation) while achieving a fixed data rate. In this work,  \emph{we focus on rate adaptive schemes where users transmit with their maximum power in order to maximize their rate depending on the existing channel and interference conditions.}
Therefore, for the scope of this paper, we assume that all users transmit with their maximum power $P_{\mathrm{max}}$ on a given subcarrier with rate adaptation depending on their channel qualities. At this point, it is also important to emphasize that this is not a limitation and the approach can be extended for various uplink power control mechanisms.  The instantaneous signal to noise ratio (SNR) $\gamma$ of each user  can then be written as follows:
\begin{equation}
\label{1}
\gamma=P_{\mathrm{max}} C \frac{{r}^{-\beta} \psi {\eta} }{\sigma^2}= \bar{K} {{r}^{-\beta} \zeta}
\end{equation}
where $\bar{K}=\frac{P_{\mathrm{max}} C}{\sigma^2}$,
$C$ is the path loss constant, $r$ is the user distance from its serving BS, $\psi$ and $\eta$ denotes the shadowing and small scale  fading coefficient between user and BS on a given subcarrier, respectively, $\beta$ is the path loss exponent, $\sigma$ denotes the thermal noise at the receiver and $\zeta$ is the composite fading. Note that all users are assumed to be associated with their closest BS \cite{TWC,novlan}, therefore $r \leq R$. 

Each cell is divided into $K$ concentric circular regions. Since path loss decays exponentially from cell center to cell edge, therefore, we consider discretization of cellular region in such a way that the path loss decay within each circular region remains constant or uniform. The main motivation for dividing the cell into a discrete set of circular regions relies on the fact that the channel statistics of the users located in a given circular region become relatively similar especially for large values of $K$.  More explicitly, the characterization of the circular regions can be demonstrated as follows:
\begin{equation}
\mathrm{log}_{10}r_k={\frac{\kappa+ 10 \beta \mathrm{log}_{10} r_{k-1}}{10 \beta}},\:\:\: r\leq R
\end{equation}
where $\kappa$ is the path loss decay within each circular region [dB]. Due to the exponential nature of the path loss, it varies rapidly near the cell center than at the cell edge, therefore, (i) each of the $k^{\mathrm{th}}$ circular region bounded by two adjacent rings, i.e., $r_k$ and $r_{k-1}$ possess non-uniform width $\Delta_k=r_k-r_{k-1}$; (ii) the number of circular regions are high in the cell center than at the cell edge; (iii)  the average number of users located within $k^{\mathrm{th}}$ circular region bounded by ring $r_k$ and $r_{k-1}$ are considered to be located at $r_k$. Note that, this is an approximation which is required for deriving the analytically tractable model of ICI and in any case it is not required for the Monte-Carlo simulations. The average number of users in each ring $k$ (for analysis) can then be given as:
\begin{equation}
u_k=\frac{U(r_{k}^2-r_{k-1}^2)}{R^2} \:\:\:\:\: k=1,2,\cdots,K
\end{equation}
It is important to note that $u_k$ can be a fraction; therefore, we round off the fractional part of users.
\subsection{Main Steps of the Proposed Framework}
In order to characterize the statistics of the  uplink and downlink ICI for generalized fading channels and  various scheduling schemes, the proposed framework mandates the following steps:
\begin{enumerate}[i)]
\item Derive the distribution $f_{r_{\mathrm{sel}}}(r)$ of the distance of the allocated user $r_{\mathrm{sel}}$ in a given cell from its serving BS based on the deployed scheduling scheme. Without loss of generality, we assume the same scheduling scheme is implemented in all cells; therefore, $f_{r_{\mathrm{sel}}}(r)$ remains the same for all cells.
\item Derive the distribution $f_{\tilde{r}_{\mathrm{sel}}}(\tilde{r})$ of the distance  between the allocated user in a neighboring interfering cell and the BS of the cell of interest $\tilde{r}_{\mathrm{sel}}$.
\item Derive the distribution $f_{X_l}(x)$ of the interference from the neighboring cell $l$.
Finally, derive the MGF of the cumulative ICI, i.e., ${Y}=\sum_{l=1}^{L}X_l$.
\end{enumerate}

\section{Part I: Distribution of the Scheduled User Location}
Considering the high dependence of the uplink ICI on the location of the scheduled users in the neighbor interfering cells which  in turn depends on the deployed scheduling schemes,  we derive in this section the distribution of the distance between the scheduled user and its serving BS in a given cell (i.e., the probability mass function (PMF) of $r_{\mathrm{sel}}$) considering the following five scheduling algorithms: greedy scheduling, proportional fair scheduling, round robin scheduling, location based round robin scheduling, and greedy round robin scheduling.
\subsection{Greedy Scheduling Scheme}
\noindent  Greedy scheduling is an opportunistic scheme that aims at maximizing the network throughput by taking full advantage of multiuser diversity. However, it suffers from low fairness among users which makes it less attractive for network operators.  The procedure for determining the PMF of $r_{\mathrm{sel}}$ considering greedy scheduling is divided into two steps:\\
\noindent\textbf{Step~1~(\rm\textit{Selecting the user with the highest SNR in ring $k$}):}
Since the path loss decay within each circular region is considered to be uniform, we approximate the distance of all  users located within $k^{\mathrm{th}}$ circular region by ring $r_k$ for analytical tractability as we already mentioned in Section II. In this step, we select a user with maximum SNR in each ring $k$ which  posses $u_k$ users. Thus, selecting a user in a ring $k$ is equivalent to selecting the user with maximum channel gain among all the users in ring $k$, i.e.,
\begin{equation}
\zeta_k=\text{max}\{\zeta_1,\zeta_2,\cdots,\zeta_i,\cdots, \zeta_{u_k}\}
\end{equation}
where $\zeta_i$ is the composite fading channel gain  between user $i$ and its BS on a given subcarrier. 
The CDF and PDF of the maximum channel gain $\zeta_k$ can be written as follows:
\begin{equation}
\label{CDF1}
F_{\zeta_k}(\zeta_k)=\prod_{i=1}^{u_k}F_{\zeta_i}(\zeta_k)\stackrel{\mathrm{i.i.d}}{=}\left(F_{\zeta}(\zeta_k)\right)^{u_k}
\end{equation}
\begin{equation}
\label{PDF1}
f_{\zeta_k}(\zeta_k)=\sum_{j=1}^{u_k} f_{\zeta_j}(\zeta_k)\prod_{i=1,i\neq j}^{u_k}F_{\zeta_i}(\zeta_k)\stackrel{\mathrm{i.i.d}}{=}{u_k}f_{\zeta}(\zeta_k)\left(F_{\zeta}(\zeta_k)\right)^{u_k-1}
\end{equation}
To consider path loss, we now perform a transformation of RVs using \eqref{1}, $\gamma_{k}=\bar{K}r_{k}^{-\beta}\zeta_k$, where, $\gamma_{k}$ is the selected user SNR in each ring $k$. The CDF and PDF of $\gamma_{k}$ can then be written as follows:
\begin{equation}
\label{CDFstep1}
F_{\gamma_{k}}(\gamma_{k})= \prod_{i=1}^{u_k}F_{\zeta_i}(\bar{K}^{-1}\gamma_{k}r_{k}^{\beta} ) \stackrel{\mathrm{i.i.d}}{=}\left(F_{\zeta}(\gamma_{k}r_{k}^{\beta} \bar{K}^{-1})\right)^{u_k}
\end{equation}
\begin{equation}
\label{PDFstep1}
f_{\gamma_{k}}(\gamma_{k})=\frac{1}{r_k^{-\beta}} \sum_{j=1}^{u_k} f_{\zeta_j}(\gamma_{k}r_{k}^{\beta}\bar{K}^{-1} )\prod_{i=1,i\neq j}^{u_k}F_{\zeta_i}(\gamma_{k}r_{k}^{\beta} \bar{K}^{-1})\stackrel{\mathrm{i.i.d}}{=}\frac{u_k}{r_{k}^{-\beta}}f_{\zeta}(\gamma_{k} r_{k}^{\beta}\bar{K}^{-1})\left(F_{\zeta}(\gamma_{k} r_{k}^{\beta}\bar{K}^{-1})\right)^{u_k-1}
\end{equation}

\noindent\textbf{Step~2~(\rm\textit{Selecting the user with maximum SNR among $K$ rings}):}
In this step, we  compute the probability of selecting the $k^\mathrm{th}$ ring among all other rings. It is important to note that this is equivalent to selecting the ring $k$ which possesses the user with the highest SNR among all rings.
Conditioning on $\gamma_{k}$, the PDF of $r_{\mathrm{sel}}$  can be written explicitly as follows:
\begin{equation}
\label{Final0}
P(r_{\mathrm{sel}}=r_{k}|\gamma_{k})=\prod_{i=1,i\neq k}^{K}p(\gamma_{i}\leq \gamma_{k})=\prod_{i=1,i\neq k}^{K} F_{\gamma_{i}}(\gamma_{k})
\end{equation}
By averaging over the distribution of $\gamma_{k}$, the final expression for the PMF of $r_{\mathrm{sel}}$ is
\begin{equation}
\label{Final}
P(r_{\mathrm{sel}}=r_k)=\int_0^\infty\left({\prod_{i=1,i\neq k}^{K}F_{\gamma_{i}}( \gamma_{k})}\right)f_{\gamma_{k}}(\gamma_{k}) d\gamma_{k}
\end{equation}
Using \eqref{CDFstep1}, \eqref{Final} can be written for i.i.d. case as follows:
\begin{equation}
\label{Final2}
P(r_{\mathrm{sel}}=r_{k})=\int_0^\infty\prod_{i=1,i\neq k}^{K}
\left(F_{\zeta}(\gamma_{k}r_{i}^{\beta})\right)^{u_i}
\frac{u_k f_{\zeta}(\gamma_{k}r_{k}^{\beta})}{r_{k}^{-\beta}}\left(F_{\zeta}(\gamma_{k} r_{k}^{\beta} )\right)^{u_k-1}
\end{equation}
where $r_{\mathrm{sel}} \in [0,R]$.
The results in \eqref{Final2} are generalized for any shadowing and fading statistics. Even though \eqref{Final2} is not a closed form expression, the integration can be solved accurately using standard mathematical software packages such as \texttt{MAPLE} and \texttt{MATHEMATICA}.

\subsection{Proportional Fair Scheduling Scheme}
The proportional fair scheduling scheme  allocates the subcarrier to the user with the largest normalized SNR (${\gamma}/{\bar{\gamma}}$)  \cite{prop1}, where $\gamma$ and $\bar{\gamma}$ denote the instantaneous SNR and the short term average SNR of a given user, respectively. In other words, the selection criterion is based on selecting a user who has maximum instantaneous SNR relative to its own average SNR.
The distribution of $r_{\mathrm{sel}}$ can be derived as follows:\\
\noindent\textbf{Step~1~(\rm\textit{Selecting the user with maximum normalized SNR in ring $k$}):}
In this step, the performance of proportional fair scheduling scheme is independent of the path loss factor if users are moving relatively slowly, i.e., their path loss remains nearly the same on a short term basis. 
In this case, the problem of selecting the maximum normalized SNR in a ring $k$ can be written as:
\begin{equation}
\label{Norm1}
{\zeta_k}
=\text{max}\left\{\frac{\zeta_1}{\bar{\zeta}_1},\frac{\zeta_2}{\bar{\zeta}_2},\cdots,\frac{\zeta_i}{\bar{\zeta}_i},\cdots,\frac{\zeta_{u_k}}{\bar{\zeta}_{u_k}}\right\}\:\:\:
\end{equation}
where $\bar{\zeta}_i=\int_0^\infty \zeta_i f_{\zeta_i}(\zeta_i) d\zeta_i$ is the average of the composite fading channel and $\zeta_k$ is the maximum normalized composite fading channel gain  in ring $k$.
For i.i.d. composite fading gains  of the users located in ring $k$, i.e.,  $\bar{\zeta}=\bar{\zeta}_1=\bar{\zeta}_2=\cdots \bar{\zeta}_{u_k}$, the problem of selecting the user with  maximum normalized channel gain reduces to selecting the user with the maximum channel gain, i.e.,
$
\zeta_k=\text{max}\left\{{\zeta_1},{\zeta_2},\cdots,{\zeta_i},\cdots,{\zeta_{u_k}}\right\}
$.
Thus, for any ring $k$, the CDF and PDF of the selected SNR  $\gamma_k=Kr_k^{-\beta}\zeta_k$ can be written as:
\begin{equation}
F_{\gamma_k}(\gamma_k)=
\prod_{i=1}^{u_k}F_{\zeta_i}(\bar{\zeta_i}\gamma_k r_k^{\beta}K^{-1})\stackrel{\mathrm{i.i.d}}{=}(F_{\zeta_i}(\bar{\zeta}\gamma_k r_k^{\beta}K^{-1}))^{u_k}
\end{equation}
\begin{equation}
f_{\gamma_k}(\gamma_k)=\frac{1}{r_k^{-\beta}} \sum_{j=1}^{u_k} f_{\zeta_j}(\bar{\zeta_j}\gamma_k r_k^{\beta}K^{-1})\prod_{i=1,i\neq j}^{u_k}F_{\zeta_i}(\bar{\zeta_i}\gamma_k r_k^{\beta}K^{-1})\stackrel{\mathrm{i.i.d}}{=}\frac{u_k}{r_k^{-\beta}} (F_{\zeta_i}(\bar{\zeta}\gamma_k r_k^{\beta}K^{-1}))^{u_k-1}f_{\zeta}(\bar{\zeta}\gamma_k r_k^{\beta}K^{-1})
\end{equation}
The short term average SNR of the selected user in ring $k$, i.e., $\bar{\gamma}_k$ can then be computed as $
\bar{\gamma}_k=\int_0^\infty \gamma_k f_{\gamma_{k}}(\gamma_k)d\gamma_k
$. Finally, the normalized selected SNR in each ring $k$ can be defined as $\xi_k=\frac{\gamma_k}{\bar{\gamma}_k}$ and performing a transformation of RVs, the CDF and PDF of $\xi_k$ can be given as 
$
F_{\xi_k}(\xi_k)=F_{\gamma_k}(\bar{\gamma}_k\xi_k)
$; and
$
f_{\xi_{k}}(\xi_k)=\frac{1}{\bar{\gamma}_k}f_{\gamma_k}(\bar{\gamma}_{k} \xi_k)
$, respectively.
\\
\noindent\textbf{Step~2~(\rm\textit{Selecting the ring $k$ with maximum normalized SNR from the $K$ rings}):} Once we characterize the PDF and CDF of $\xi_k$, the probability of selecting any ring $k$ can be written using \eqref{Final0} as follows:
\begin{equation}
\label{Finalnorm}
P(r_{\mathrm{sel}}=r_k)=\int_0^\infty\left({\prod_{i=1,i\neq k}^{K}F_{\xi_{i}}( \xi_{k})}\right)f_{\xi_{k}}(\xi_{k}) d\xi_{k}
\end{equation}

\subsection{Round Robin Scheduling Scheme}
Round robin scheduling is a non-opportunistic scheme where a user is selected randomly within a time slot. As each user has equal probability of allocation, the scheme can be referred to as a \emph{strictly fair} scheduling scheme. The round robin scheme provides maximum fairness among users and may serve as a lower bound in terms of network throughput which is useful in calibrating the performance of other scheduling schemes, however, the resulting network throughput is significantly low which makes it less attractive for practical implementations.  The PMF of the scheduled user location can then be given as:
\begin{equation}
P(r_{\mathrm{sel}}=r_{k}) = \frac{u_k}{U}
\end{equation}
\subsection{Location based Round Robin Scheduling Scheme}
Location based round robin is another non-opportunistic scheduling scheme which do not require any channel state information, however, it requires the location information of the users. Even though location based scheduling is not common in practice,
the location of each mobile user can be determined  at the BS using global positioning system (GPS) or estimate based on a power measurement of pilot signals from the surrounding beacons, e.g., using triangulation based techniques. In this regard, there are variety of techniques available in the literature which demonstrate  how the location of users can be evaluated at the BS (see\cite{loc1,loc2} and the references therein). Moreover, the users located in different circular regions can also be classified based on the long term average SNRs, i.e., by computing SNR thresholds for different distances (rings) \cite{yang} which is a common technique in fractional frequency reuse (FFR) systems to distinguish between cell-edge and cell center users.

In this context, we consider $W$ time slots during which the distance of the users from their serving BS will remain approximately the same. For simplicity, the number of time slots $W$ is set equal to $K$. 
At a given time slot $T_w$, we select any arbitrary user from a specific ring (analysis) and circular region (simulations) starting from the cell center. We continue to allocate the users by accessing the circular rings sequentially from cell center to cell edge. At this point, it is important to emphasize that all cells are considered to be time synchronous in allocating the users from particular rings, i.e., at a given time slot $T_w$ all cells are selecting the $w^{th}$ ring. Thus, the    PMF of $r_{\mathrm{sel}}$ for a given time slot $w$ denoted by $P(r_{\mathrm{sel}}=r_k^{T_{w}})$ can be given as:
\begin{equation}
P(r_{\mathrm{sel}}=r_k^{T_w})=
\begin{cases}
1, & \text{if }k = w\\
0, &  \text{else}
\end{cases}
\end{equation}
The scheme can produce relatively high capacity gains on average compared to the traditional round robin scheme. Moreover, the average fairness measure is very close to the traditional round robin. 

\subsection{Greedy Round Robin Scheduling Scheme}
Greedy round robin is an opportunistic scheduling scheme which captures the multiuser diversity  while maintaining some fairness among users. In this scheme,
we consider $W=K$ time slots during which the distance of the users from  their serving BSs remain nearly the same, however, the small scale fading gain on the considered subcarrier may vary from one time slot to the other. We select the user with maximum SNR  in each time slot $T_w$, however once a user is selected from a ring, all users located in that ring will not be scheduled for transmission for the next $K-1$ time slots.
Note that all BSs are considered to be time synchronized in terms of scheduling.
Clearly, the  probability of allocating a ring $k$ at $T_1$ can simply be given by \eqref{Final2}. However, the probability of selecting a ring $k$ at $T_2$ is a dependent event and can be derived using Bayes theorem as follows:
\begin{equation}
P(r_{\mathrm{sel}}=r_k^{T_{2}})=\sum_{\substack{
j=1\\
j \neq k}}^K P\left(\frac{{r_{\mathrm{sel}}=r_k}^{T_2}}{{r_{\mathrm{sel}}=r_j}^{T_1}}\right)P(r_{\mathrm{sel}}=r_j^{T_1})
\end{equation}
where,
\begin{equation}
P\left(\frac{{r_{\mathrm{sel}}=r_k}^{T_2}}{{r_{\mathrm{sel}}=r_j}^{T_1}}\right)=
\int_0^\infty\prod_{\substack{i\neq k\\ i\neq j}}^{K}\left(F_{\zeta}(\gamma_{k}r_{i}^{\beta})\right)^{u_i}\frac{u_k f_{\zeta}(\gamma_{k}r_{k}^{\beta})}{r_{k}^{-\beta}}\left(F_{\zeta}(\gamma_{k} r_{k}^{\beta} )\right)^{u_k-1}d\gamma_{k}
\end{equation}
Since the probability of allocating any ring $k$ within time slot $T_w$ depends on all previous states, therefore, the principle of Markov chain transition probabilities is not directly applicable. For more clarity,  the  probability of selecting a ring $k$ at $T_3$ is given as follows:
\begin{equation}
P(r_{\mathrm{sel}}=r_k^{T_{3}})=\sum_{m \neq k}^{K}\sum_{\substack{j \neq m\\ j\neq k}}^K P\left(\frac{{r_{\mathrm{sel}}=r_k}^{T_3}}{{r_{\mathrm{sel}}=r_j}^{T_2}\cap {r_{\mathrm{sel}}=r_m}^{T_1}}\right)P\left(\frac{{r_{\mathrm{sel}}=r_j}^{T_2}}{{r_{\mathrm{sel}}=r_m}^{T_1}}\right)P(r_{\mathrm{sel}}=r_m^{T_1})
\end{equation}
where,
\begin{equation}
P\left(\frac{{r_{\mathrm{sel}}=r_k}^{T_3}}{{r_{\mathrm{sel}}=r_j}^{T_2}\cap {r_{\mathrm{sel}}=r_m}^{T_1}}\right)=
\int_0^\infty\prod_{\substack{i\neq k \\ i\neq j, i\neq m}}^{K}\left(F_{\zeta}(\gamma_{k}r_{i}^{\beta})\right)^{u_i}\frac{u_k f_{\zeta}(\gamma_{k}r_{k}^{\beta})}{r_{k}^{-\beta}}\left(F_{\zeta}(\gamma_{k} r_{k}^{\beta} )\right)^{u_k-1}d\gamma_{k}
\end{equation}
In general, the probability of selecting any ring $k$ at a time slot $T_w$, i.e.,  $P(r_{\mathrm{sel}}=r_k^{T_{w}})$ can be written as:
\begin{multline}
\sum_{n \neq k}^{K}\sum_{\substack{s\neq n, \\s\neq k}}^K \cdots \sum_{\substack{j\neq s,n,..\\j\neq k}}
P\left(\frac{{r_{\mathrm{sel}}=r_k}^{T_w}}{{r_{\mathrm{sel}}=r_j}^{T_{w-1}}\cdots \cap {r_{\mathrm{sel}}=r_s}^{T_2}\cap {r_{\mathrm{sel}}=r_n}^{T_1} }\right) \cdots
P\left(\frac{{r_{\mathrm{sel}}=r_s}^{T_2}}{{r_{\mathrm{sel}}=r_n}^{T_1}}\right)P(r_{\mathrm{sel}}=r_n^{T_1})
\end{multline}
{\it Computational Efficiency:} The time complexity of the greedy round robin scheme is  heavily based on the computational time of the  \texttt{NIntegrate} operation in \texttt{MATHEMATICA}. One  \texttt{NIntegrate} operation requires around 0.95~sec which is equivalent to the computational complexity of (i) \emph{greedy} scheme and (ii) the \emph{first time slot of the greedy round robin} scheme.  Monte-Carlo simulation time required for 100,000 trials in \emph{greedy} scheme requires around 150.671928~sec which is more than 150 times the computational complexity of \texttt{NIntegrate} operation. This fact demonstrates the computational efficiency of \emph{greedy} scheme in comparison to Monte-Carlo simulations. 
However, in the second  time slot, greedy round robin scheme requires $K-1$ integrations whereas, for the third time slot $K-2$ integrations are required. Therefore, in general the computational complexity of greedy round robin scheme at any time slot $w$ can be given as:
\begin{equation}
{\rm Computational \:time}[\mathrm{sec}]=0.95+\sum_{i=2}^{w} 0.95(K-i+1);\:\:\:w\leq K
\end{equation}
where $K$ denotes the number of rings. Therefore considering  $w=15$ and $K=w$ for greedy round robin,  the analytical time complexity  is around 113~sec which is still lower than the Monte-Carlo simulation time required for the \emph{greedy} scheme. Therefore, even though the greedy round robin scheme is computationally complex for large  $W$, the evaluation time remains comparable to the Monte-Carlo simulations.

\subsection{Evaluating the Joint PMF of $r_{\mathrm{sel}}$ and $\theta$}
Note that, $P(r_{\mathrm{sel}}=r_{k})$ derived for all of the above scheduling schemes is the marginal PMF of $P(r_{\mathrm{sel}}=r_k,\mathcal{\theta}=\theta_i)$ where $\mathcal{\theta}$ denotes the angle of the allocated user with respect to the serving BS and it is uniformly distributed from $0$ to $2\pi$ (see Fig. 1). Although the PDF of $\mathcal{\theta}$ is continuous we can discretize it for analytical consistency and complexity reduction.
Consider discretizing the range of RV $\theta$ in ${\mathcal{I}}$ uniform angular intervals of desired accuracy. Thus $P(\mathcal{\theta}=\theta_i)=\frac{1}{{\mathcal{I}}}$ where $\theta_i$ denotes any discrete value that the RV $\mathcal{\theta}$ can take. Since $r_{\mathrm{sel}}$ and $\mathcal{\theta}$ are independent, their joint PMF can be written as follows:
\begin{equation}
\label{IntPMF}
\begin{split}
P(r_{\mathrm{sel}}=r_k, \mathcal{\theta}=\theta_i)=P(r_{\mathrm{sel}}=r_k)P(\mathcal{\theta}=\theta_i)=\frac{P(r_{\mathrm{sel}}=r_k)}{\mathcal{I}}
\end{split}
\end{equation}

\section{Part I: Distribution of Intercell Interference from One Cell}
\begin{figure}[t]
  \centering
\includegraphics[width=0.8\columnwidth,keepaspectratio=true]{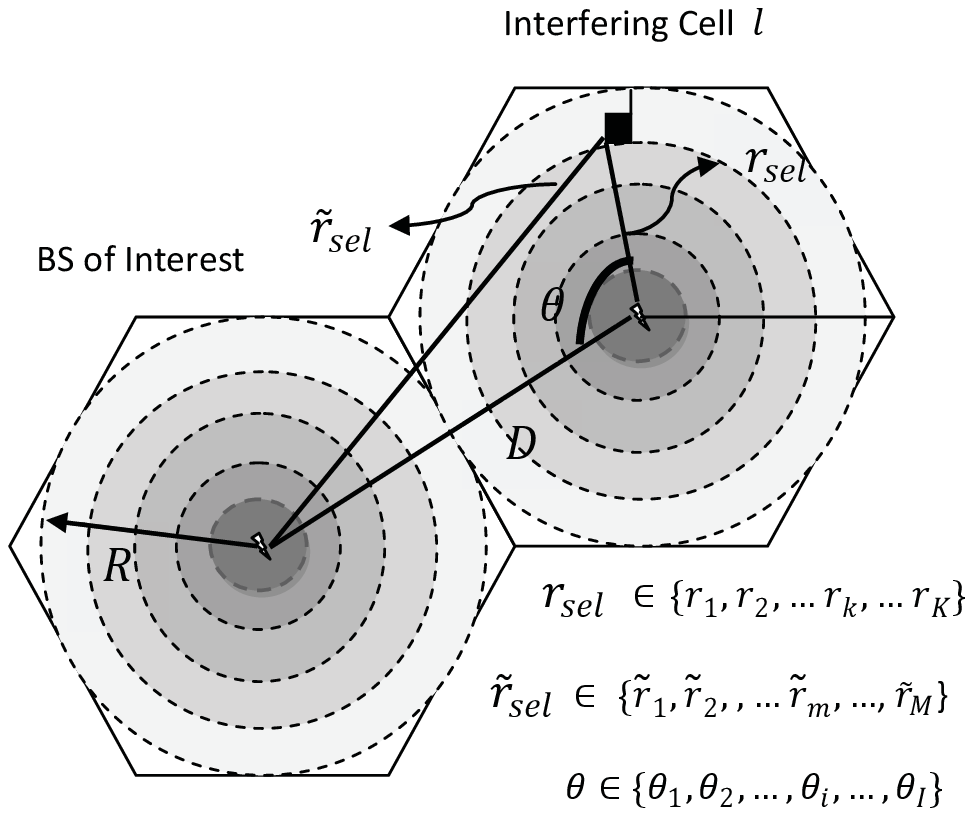}
\caption{Geometrical illustration of dividing the cellular network into multiple rings of non-uniform width $\Delta_k$.}
\label{Cap0}
\end{figure}

The derivation for the distribution of the ICI from an interfering cell $l$, i.e., $f_{X_l}(x)$, depends on the distribution of the distance between the allocated user in the interfering cell $l$ and the BS of interest, i.e., $f_{\tilde{r}_{sel_l}}(\tilde{r})$.  As mentioned earlier, each interfering cell is assumed to have identical conditions in a given time slot. Therefore, $f_{\tilde{r}_{\mathrm{sel}}}(\tilde{r})$ applies to all interfering cells and, thus, we will drop the subscript $l$ in the sequel to simplify notation. Using the cosine law (see \figref{Cap0}), we can write:
\begin{equation}
\label{cosine law}
\begin{split}
\tilde{r}_{{\mathrm{sel}}}^2=r_{{\mathrm{sel}}}^2+D^2-2{r}_{{\mathrm{sel}}}D \;\mathrm{cos}{\theta}
\end{split}
\end{equation}
$\tilde{r}_{{\mathrm{sel}}}$ is the distance of the allocated user in the interfering cell $l$ from the BS of interest, $r_{{\mathrm{sel}}}$ is the distance of the allocated user from its serving BS, i.e., (BS $l$), $\theta \in [0, 2\pi]$ and $D = 2R$ since we consider universal frequency reuse with one tier of interfering cells. The approach can be extended to any number of tiers in a straightforward manner.
In order to determine the PMF of $\tilde{r}_{\mathrm{sel}}$ where $ \tilde{r}_{\mathrm{sel}} \in [D-R, D+R] $, first of all we compute $\tilde{r}_{i,k}$ for given $\theta_i$ and $r_k$ using \eqref{cosine law} as follows:
\begin{equation}
\begin{split}
\tilde{r}_{i,k}^2=r_{k}^2+D^2-2{r_k}D \;\mathrm{cos}{\theta_i} \:\:\: \forall r_k , \forall \theta_i
\end{split}
\end{equation}
where, $\tilde{r}_{i,k}$ denotes the interfering distance from a specified polar coordinate $( r_k,\theta_i)$ in the interfering cell to the BS of interest located at a distance $D$ (see \figref{Cap0}).
In addition, it is worth to mention that $\tilde{r}_{i,k}$ are the points at which $P(\tilde{r}_{\mathrm{sel}}=\tilde{r}_{i,k})$ can be defined using \eqref{IntPMF} as follows:
\begin{equation}
P(\tilde{r}_{\mathrm{sel}}=\tilde{r}_{i,k})\\=\frac{P(r_{\mathrm{sel}}=r_k)}{\mathcal{I}}
\end{equation}
The two dimensional data set of $\tilde{r}_{\mathrm{sel}}$, at which $P(\tilde{r}_{\mathrm{sel}}=\tilde{r}_{i,k})$ is defined, can then be grouped into $M$ segments of any arbitrary width $\Delta$. This can be done by dividing the distance between $D-R$ and $D+R$ into $M$ equal segments of width $\Delta$ and mapping $\tilde{r}_{i,k}$ accordingly. Clearly, by adding all the probabilities for which $\tilde{r}_{\mathrm{sel}}$ lies in the $m^{\mathrm{th}}$ segment we get the probability of $\tilde{r}_{\mathrm{sel}}=\tilde{r}_m$:
\begin{equation}
\label{step}
P(\tilde{r}_{\mathrm{sel}}=\tilde{r}_m)= \sum_{\tilde{r}_{i,k} \in[\tilde{r}_m - \frac{\Delta}{2}, {\tilde{r}_m+\frac{\Delta}{2}}]}
P(\tilde{r}_{\mathrm{sel}}=\tilde{r}_{i,k})
\end{equation}
where $\tilde{r}_m$ denotes any discrete value that the RV $\tilde{r}_{\mathrm{sel}}$ can take. Recall $X=\bar{K}\tilde{r}^{-\beta} \chi$, therefore the PDF of $X$ conditioned on $\tilde{r}_{\mathrm{sel}}$ can be determined by RV transformation as follows:
\begin{equation}
\label{expo}
f_{X|\tilde{r}_{\mathrm{sel}}}=\frac{f_\chi(x\tilde{r}_{\mathrm{sel}}^{\beta}\bar{K}^{-1})}{\bar{K}\tilde{r}_{\mathrm{sel}}^{-\beta}}
\end{equation}
Averaging over the PMF of $\tilde{r}_{\mathrm{sel}}$, the distribution of the ICI, $f_X(x)$, from any cell $l$ can be given as:
\begin{equation}
\label{ICI}
f_{X}(x)=\sum_{\tilde{r}_m=\tilde{r}_1}^{\tilde{r}_M} \frac{f_\chi(x\tilde{r}_m^{\beta}\bar{K}^{-1})}{\bar{K}\tilde{r}_m^{-\beta}}P(\tilde{r}_{\mathrm{sel}}=\tilde{r}_m)
\end{equation}
It is important to emphasize that the derivation of the distribution of ICI is based on the scheduling decisions of interfering cells at a given time slot. Therefore, the parameter $\tilde{r}_m$ of the ICI distribution varies from one time slot to the other for the location based round robin and greedy round robin schemes.

\section{Part I: MGF of the Cumulative Uplink and Downlink Intercell Interference }
Computing the distribution of the cumulative ICI $Y$  requires the convolution of the PDF of $L$ RVs $X_l$, $\forall l=1,2,\cdots L$, which is a tedious task for many practical scenarios. To avoid the convolutions, we utilize an MGF based approach and derive the expression for the MGF of the cumulative ICI $Y$. 
\subsection{Derivation for the Uplink Interference}
Since each cell is considered to have same scheduling scheme deployed, therefore,  the MGF of the cumulative interference considering  i.i.d. interferers can be calculated as follows:
\begin{equation}
\mathcal{M}_{Y}(t)=\prod_{l=1}^{L} \mathcal{M}_{X_l}(t)=\left(\mathcal{M}_{X}(t)\right)^L=\left(\mathbb{E}[e^{tx}]\right)^L
\end{equation}
Looking at the structure of \eqref{ICI}, we can write $\mathcal{M}_{X}(t)$ as:
\begin{equation}\label{cumMGF}
\begin{split}
\mathcal{M}_{X}(t)=\int_0^\infty e^{tx}f_{X}(x) dx=\mathcal{M}_{X}(t)=\sum_{\tilde{r}_m=\tilde{r}_1}^{\tilde{r}_M}
\frac{P(\tilde{r}_{\mathrm{sel}}=\tilde{r}_m)}{\bar{K}\tilde{r}_m^{-\beta}} \int_0^\infty  e^{tx}{f_\chi(x\tilde{r}_m^{\beta}\bar{K}^{-1})}dx
\end{split}
\end{equation}
The derived expression is generic and applies to any composite fading distribution. Next, we will present explicit MGF expressions for the uplink ICI considering three typically used practical fading models.\\
{\it \bf Special Case 1: Rayleigh fading -$\zeta,  \chi \sim \mathrm{Exp}(\lambda)$:}
In this case, the small scale fading coefficient on a given subcarrier is considered to be Rayleigh distributed whereas the effect of shadowing is not considered. The distribution of interference considering a single interfering cell can then be derived as follows:
\begin{equation}
\label{HEdist}
f_{X}(x) = \sum_{\tilde{r}_m=\tilde{r}_1}^{\tilde{r}_M}\bar{K}^{-1}\lambda\tilde{r}_m^{\beta} e^{-\lambda \tilde{r}_m^\beta \bar{K}^{-1} x} P(\tilde{r}_{\mathrm{sel}}=\tilde{r}_m)
\end{equation}
Note that \eqref{HEdist} is a Hyper-Exponential distribution with parameter $\bar{K}^{-1} \lambda \tilde{r}_m^\beta$. Thus, using the MGF of the Hyper-Exponential distribution, $\mathcal{M}_{Y}(t)$ can be derived as follows:
\begin{equation}
\label{finalmgfHE}
\mathcal{M}_{Y}(t)=\left(\sum_{\tilde{r}_m=\tilde{r}_1}^{\tilde{r}_M}\frac{\bar{K}^{-1}\lambda \tilde{r}_m^\beta}{\bar{K}^{-1}\lambda \tilde{r}_m^\beta-t}P(\tilde{r}_{\mathrm{sel}}=\tilde{r}_m)\right)^L
\end{equation}
{\it  \bf Special Case 2: Generalized-$\mathcal{K}$ composite fading -{$\zeta,  \chi \sim \mathcal{K}_G(m_s,m_c, \Omega)$:}}
In wireless channels, shadowing and fading across the channel between a user and BS can be jointly modeled by a composite fading distribution.
A closed form composite fading model, namely Generalized-$\mathcal{K}$ also referred to as Gamma-Gamma distribution, has been recently introduced in \cite{KG} which is general enough to model well-known shadowing and fading distributions such as log-normal, Nakagami-\emph{m}, Rayleigh etc.  Using \eqref{ICI}, $f_{X}(x)$ in this case can be derived as follows:
\begin{equation}
f_{X}(x) = \sum_{\tilde{r}_m=\tilde{r}_1}^{\tilde{r}_M}  \frac{2(x{\tilde{r}_m}^{\beta}\bar{K}^{-1})^{\frac{m_c+m_s-2}{2}}}{\bar{K}\tilde{r}_m^{-\beta}\Gamma(m_c)\Gamma(m_s)} \mathbb{K}_{m_s-m_c}\left(b\sqrt{x\tilde{r}_m^\beta \bar{K}^{-1}}\right)\left(\frac{b}{2}\right)^{m_c+m_s} P(\tilde{r}_{sel}=\tilde{r}_m)
\end{equation}
where, $\mathbb{K}_v(.)$ denotes the modified Bessel function of second kind with order $v$, $b=2\sqrt{\frac{m_c m_s}{\Omega}}$. Performing some algebraic manipulations and using  \cite[Eq. 6.643/3]{book}, the expression for $\mathcal{M}_{X}(t)$ can be derived as:
\begin{equation}
\mathcal{M}_{X}(t)=\sum_{\tilde{r}_m=\tilde{r}_1}^{\tilde{r}_M}
P(\tilde{r}_{sel}={\tilde{r}_m}^\beta)\mathbb{W}_{\frac{1-m_c-m_s}{2},\frac{m_c-m_s}{2}}\left(-\frac{\tilde{r}_m^{\beta}b^2}{4\bar{K}t}\right)
e^{\frac{b^2\tilde{r}_m^{\beta}}{8\bar{K}t}}
\left(\frac{-b^2\tilde{r}_m^{\beta}}{4\bar{K}t}\right)^{(\frac{m_s+m_c-1}{2})}
\end{equation}
where, $\mathbb{W}$ denotes the Whittaker function.
Finally, $\mathcal{M}_{Y}(t)$ can be written as follows:
\begin{equation}
\label{finalmgfHG1}
\mathcal{M}_{Y}(t)=\left(\sum_{\tilde{r}_m=\tilde{r}_1}^{\tilde{r}_M}
P(\tilde{r}_{sel}={\tilde{r}_m}^\beta)\mathbb{W}_{\frac{1-m_c-m_s}{2},\frac{m_c-m_s}{2}}\left(-\frac{\tilde{r}_m^{\beta}b^2}{4\bar{K}t}\right)
e^{\frac{b^2\tilde{r}_m^{\beta}}{8\bar{K}t}}
\left(\frac{-b^2\tilde{r}_m^{\beta}}{4\bar{K}t}\right)^{(\frac{m_s+m_c-1}{2})}\right)^L
\end{equation}
Integrating the CDF and MGF of Generalized-$\mathcal{K}$ RV which involves Meijer-G and Whittaker functions, respectively, in \texttt{MATHEMATICA} and \texttt{MAPLE} can be time consuming. 
\\
{\it \bf Special Case 3: Gamma Composite Fading -{$\zeta, \chi \sim \mathrm{Gamma}(m_s,m_c)$}}:
In \cite{KG}, the authors proposed an accurate approximation of the Generalized-$\mathcal{K}$ RV by the more tractable gamma distribution  using  moment matching method.
The approximation  provides a simplifying model for the composite fading in wireless communication systems.
Using \eqref{ICI}, $f_{X}(x)$ can be written in this case as:
\begin{equation}
f_{X}(x) = \sum_{\tilde{r}_m=\tilde{r}_1}^{\tilde{r}_M} \frac{e^{-\frac{x \tilde{r}_m^\beta \bar{K}^{-1}}{m_c}}(x \tilde{r}_m^\beta \bar{K}^{-1})^{m_s-1}}{\bar{K} \tilde{r}_m^{-\beta} \Gamma(m_s) m_c^{m_s}} P(\tilde{r}_{sel}=\tilde{r}_m)
\end{equation}
Performing some algebraic manipulations and letting $y=x(\frac{\tilde{r}_m^{\beta}}{m_c}-t)$, $\mathcal{M}_{X}(t)$ can be derived as follows:
\begin{equation}
\mathcal{M}_{X}(t)=\sum_{\tilde{r}_m=\tilde{r}_1}^{\tilde{r}_M}  \frac{P(\tilde{r}_{sel}=\tilde{r}_m)(\bar{K}^{-1}\tilde{r}_m^{\beta})^{m_s-1}}{\bar{K}\tilde{r}_m^{-\beta}\Gamma(m_s) \left(m_c (\frac{\tilde{r}_m^{\beta}\bar{K}^{-1}}{m_c}-t)\right)^{m_s}}
\int_0^\infty  e^{-y}y^{m_s-1}dy=\sum_{\tilde{r}_m=\tilde{r}_1}^{\tilde{r}_M}  P(\tilde{r}_{sel}=\tilde{r}_m)\left(\frac{\bar{K}^{-1}\tilde{r}_m^{\beta}}{\bar{K}^{-1}\tilde{r}_m^{\beta} -m_ct}\right)^{m_s}
\end{equation}
Finally, $\mathcal{M}_{Y}(t)$ can be written as follows:
\begin{equation}
\label{finalmgfHG2}
\mathcal{M}_{Y}(t)=\left(\sum_{\tilde{r}_m=\tilde{r}_1}^{\tilde{r}_M} P(\tilde{r}_{\mathrm{sel}}=\tilde{r}_m)\left(\frac{\bar{K}^{-1}\tilde{r}_m^{\beta}}{\bar{K}^{-1}\tilde{r}_m^{\beta} -m_ct}\right)^{m_s}\right)^L
\end{equation}
\subsection{ Derivation for the  Downlink Interference:}
The downlink interference $X_l$ considering a single interfering cell $l$ can be written as:
\begin{equation}
X_l=  \bar{r}_l^{-\beta} \chi_l
\end{equation}
where $\chi_l$ is the interfering statistics from the $l^{\mathrm{th}}$ neighboring BS, and $\bar{r}_l$ is the distance of $l^{\mathrm{th}}$ interfering BS from the scheduled  mobile receiver. 
\emph{At this point, it is important to highlight that for a given  $r_k$ and $\theta_i$ of a scheduled mobile user in the cell of interest, the distance of all interfering BSs can be calculated using cosine law which is not the same as in the uplink, where all $\tilde{r}_l$ are independent from one another}. Therefore, conditioned on the location of the mobile receiver, the distribution of the downlink cumulative  ICI, i.e., $Y|{r_k,\theta_i}$, is simply a weighted sum of the distribution of the interfering channel statistics $\chi$. More precisely and by using the symmetry of the grid model the PDF of the cumulative ICI $(Y=\sum_{l=1}^L X_l$) can be given as:
\begin{equation}
f_Y(y)=\sum_{{r}_{k}=r_1}^{{r}_K} \sum_{{\theta}_{i}=0} ^{2\pi }P({r}_{sel}={r}_{i,k})f_{Y|{r_k,\theta_i}}(y|{r_k,\theta_i})
\end{equation}
where
\begin{equation}
Y|{r_k, \theta_i}= \sum_{l=1}^{L}\bar{r}_l^{-\beta}\chi_l
\end{equation}
and the distance of all interferers can be determined using cosine law as follows: 
\begin{equation}
\bar{r}_l=\sqrt{D^2+r_k^2-2r_k D \mathrm{cos}({(l-1)\pi/3+\pi/6-\theta_i}}) \:\: \forall l=1,...L
\end{equation}
Given the distance of the interferers, the conditional MGF of the cumulative interference  can be calculated as follows:
\begin{equation}
\mathcal{M}_{Y|{r_k, \theta_i}}(t)= \prod_{l=1}^{L} \mathcal{M}_{\chi}(\bar{r}_l^{-\beta}t)
\end{equation}
Finally the MGF of the cumulative ICI can be calculated as follows:
\begin{equation}\label{downlink}
\mathcal{M}_{Y}(t)= \sum_{{r}_{k}=r_1}^{{r}_K} \sum_{{\theta}_{i}=0} ^{2\pi }P({r}_{sel}={r}_{i,k})\prod_{l=1}^{L} \mathcal{M}_{\chi}(\bar{r}_l^{-\beta}t)
\end{equation}
The expression for the MGF of the cumulative ICI in \eqref{downlink} is general for any kind of composite channel fading models. The explicit expressions for three above discussed practical fading models can also be obtained in a straightforward manner.

\section{Part I: Evaluation of Important Network Performance Metrics}
In this section, we demonstrate the significance of the derived  MGF expressions of the cumulative ICI  in quantifying important network performance metrics such as the outage probability $P_{\mathrm{out}}$, ergodic capacity $\mathcal{C}$ and average fairness $\mathcal{F}$ among users numerically.

\vspace{3mm}
\noindent
{\it \bf Evaluation of Outage Probability:}
The outage probability is typically defined as the probability of the instantaneous interference-to-signal-ratio to exceed a certain threshold. In order to evaluate $P_{\mathrm{out}}$ numerically, we use the MGF based technique introduced in  \cite{outage} for interference limited systems.
Firstly, we define a new RV, ${Z}=q\sum_{l=1}^{L} X_l-X_0=qY-X_0$, where $q$ is the outage threshold and $X_0$ is the corresponding signal power  of the scheduled user  in the central cell. An outage event occurs when $p({Z} \geq 0)$, i.e., when the interference exceeds the corresponding signal power. This decision problem is solved in \cite{outage} by combining the characteristic function of $Z$ and residue theorem. The characteristic function of ${Z}$ is defined as $\phi_{Z}(j\omega)=\mathbb{E}[e^{j{Z}\omega}]$. Considering interference $Y$ and signal power $X_0$ to be independent, the expression for $\phi_{Z}(j\omega)$ can be given as,
$
\phi_{Z}(j\omega)=\phi_{Y}(jq\omega)\phi_{X_0}(-j\omega)
$; where $\phi_Y(q j\omega)$ can be given by \eqref{finalmgfHE}, \eqref{finalmgfHG1}, and \eqref{finalmgfHG2} for different fading models. In general, the characteristic function of $X_0$ can be calculated as follows \cite{yilmaz}:
\begin{equation}\label{xo}
\phi_{X_0}(\omega)=\mathbb{E}(e^{j\omega x_0})=\int_0^\infty e^{j \omega x_0} f_{X_0}(x_0) d x_0=j \omega\int_0^\infty e^{j \omega x_0} F_{X_0}(x_0) d x_0
\end{equation}
where $F_{X_0}(x_0)=\prod_{i=1}^K F_{\gamma_k}(x_0)$ for opportunistic schemes and
compact closed form expressions  of $\phi_{X_0}(j \omega)$ are available in the literature \cite[Eq. 19]{romero}.
For non-opportunistic scheduling schemes 
$\phi_{X_0}(\omega)= \sum_{{r}_k={r}_1}^{{r}_K}  \phi_{\zeta|r_k}(j \omega) P(r_{\mathrm{sel}}={r}_k)$, where $\phi_{\zeta|r_k}(j \omega)$ is the characteristic function of $\zeta$ in ring $k$.  The outage probability can then be computed by using the  classical lemma introduced in \cite{outage} as follows:
\begin{equation}
\label{QTZhang}
P_{\mathrm{out}}=\frac{1}{2} +\frac{1}{\pi }\int_{0}^\infty \mathrm{Im}\left(\frac{\phi_Z(\omega)}{\omega}\right) d\omega
\end{equation}
where $\mathrm{Im}(\phi_Z(\omega))$ denotes the imaginary part of $\phi_Z(\omega)$. 
Using \eqref{QTZhang}, the outage probability can be evaluated using any standard mathematical software packages such as \texttt{MATHEMATICA}. 

\vspace{3mm}\noindent
{\it \bf Evaluation of Ergodic Network Capacity:}
Another important performance evaluation parameter is the network ergodic capacity $\mathcal{C}$, i.e.,
\begin{equation}
\label{capdef}
\mathcal{C}=\mathbb{E}\left[\mathrm{log_2}\left(1+\frac{X_0}{\sum_{l=1}^L{X}_l+\sigma^2}\right)\right]
\end{equation}
Usually, the computation of \eqref{capdef} requires $(L+1)$-fold numerical integrations. To avoid this, we utilize the efficient lemma derived in \cite{hamdi} with a slight modification to take thermal noise into account and compute $\mathcal{C}$ as follows:
\begin{equation}
\label{lemma}
\mathbb{E}\left[\mathrm{ln}\left(1+\frac{X_0}{\sum_{l=1}^L{X}_l+\sigma^2}\right)\right]=\int_0^\infty \frac{\mathcal{M}_Y(t)-\mathcal{M}_{X_0,Y}(t)}{t} e^{-\sigma^2 t} dt
\end{equation}
where, $\mathcal{M}_Y(t)=\mathbb{E}[e^{-t\sum_{l=1}^{L} X_l}]$ and $\mathcal{M}_{X_0,Y}(t)=\mathbb{E}[e^{-t(X_0+\sum_{l=1}^{L} X_l)}]=\mathbb{E}[e^{-t(X_0+Y)}]$. Note that this is the definition of MGF as defined in \cite{hamdi} which is not the same as our definition. Thus,
we can use $\mathcal{M}_Y(t)$ from \eqref{finalmgfHE}, \eqref{finalmgfHG1} and \eqref{finalmgfHG2} directly with a sign change of $jw$.
Moreover, \eqref{capdef} can also be solved efficiently  by expressing it in terms of the weights and abscissas of a Laguerre orthogonal polynomial \cite{hamdi}:
\begin{equation}
\label{lemma1}
\mathbb{E}\left[\mathrm{ln}\left(1+\frac{X_0}{\sum_{l=1}^L{X}_l+1}\right)\right]=
\sum_{\epsilon =1}^E \alpha_\epsilon \frac{\mathcal{M}_Y(\xi_\epsilon)-\mathcal{M}_{X_0,Y}(\xi_\epsilon)}{\xi_\epsilon} +R_E
\end{equation}
where $\xi_\epsilon$ and $\alpha_\epsilon$ are the sample points and the weight factors of the Laguerre polynomial, tabulated in \cite{book}, and $R_E$ is the remainder. Note that the MGF of $X_0$ can be calculated as explained in \eqref{xo}.
\\
{\it \bf Evaluation of Average Fairness:}
In order to quantify the degree of fairness among different scheduling schemes, we use the notion developed in \cite{elliott}. The average fairness of a scheduling scheme with $U$ users can be given as,
$
\mathcal{F}=-\sum_{i=1}^{U} p_i\frac{ \mathrm{log}_{10} p_i}{\mathrm{log}_{10}U}
$,
where $p_i$ is the proportion of resources allocated to a user $i$ or the access probability of user $i$. A system is strictly fair if each user has equal probability to access the channel and in such case the average fairness becomes one. The other extreme occurs when the channel access is dominated by a single user; in such case, the average fairness reduces to zero. The average fairness can be easily computed using our derived results as follows:
\begin{equation}
\mathcal{F}=-\sum_{k=1}^{K} P(r_{\mathrm{sel}}=r_k) \frac{ \mathrm{log}_{10} P(r_{\mathrm{sel}}=r_k)- \mathrm{log}_{10} {u_k}}{\mathrm{log}_{10}U}
\end{equation}
where $u_k$ denotes the number of users in a ring $k$.
\section{Part I: Numerical and Simulation Results}
In this section, we first define the system parameters and describe the Monte-Carlo simulation setup which is required to demonstrate the accuracy of the derived expressions.  
We then address some important insights and study the performance trends of  different scheduling schemes.
\subsection
{ Parameter Settings and Simulation Setup:}
The radius $R$ of the cell is set to 500m and the cell is decomposed into non-uniform circular regions of width $\Delta_k$. The path loss variation within each circular region is set to $\kappa=2$dB.
For each Monte-Carlo simulation trial, we generate $U$ uniformly distributed users in a circular cell of radius $R$. Each user has instantaneous SNR given by \eqref{1} and short term average SNR $(\bar{\gamma})$.   We allocate a user with maximum instantaneous SNR in the greedy scheme whereas in the proportional fair scheme we allocate a user based on the maximum normalized SNR. For the round robin scheme, we select any user arbitrarily.  For location based round robin  we select a user randomly from the $w^{th}$ ring in a time slot $w$ whereas we select a user with maximum SNR  without considering  the users of the previously allocated $w-1$ rings for the greedy round robin scheme. 
Next, we calculate the distances of the selected users, i.e., $r_{\mathrm{sel}}$, from their serving BS and compute the distances to the BS of interest, i.e., $\tilde{r}_{\mathrm{sel}}$ for all scheduling schemes. 
The process repeats for large number of Monte-Carlo  simulation trials. The  distance data is then analyzed by creating a histogram whose bins are given by $[1,\cdots r_{k-1}, r_k,  r_{k+1}, \cdots R]$. 
\subsection
{ Results and Discussions:}
\begin{figure}[t]
  \centering
  \includegraphics[totalheight=5in,width=6in]{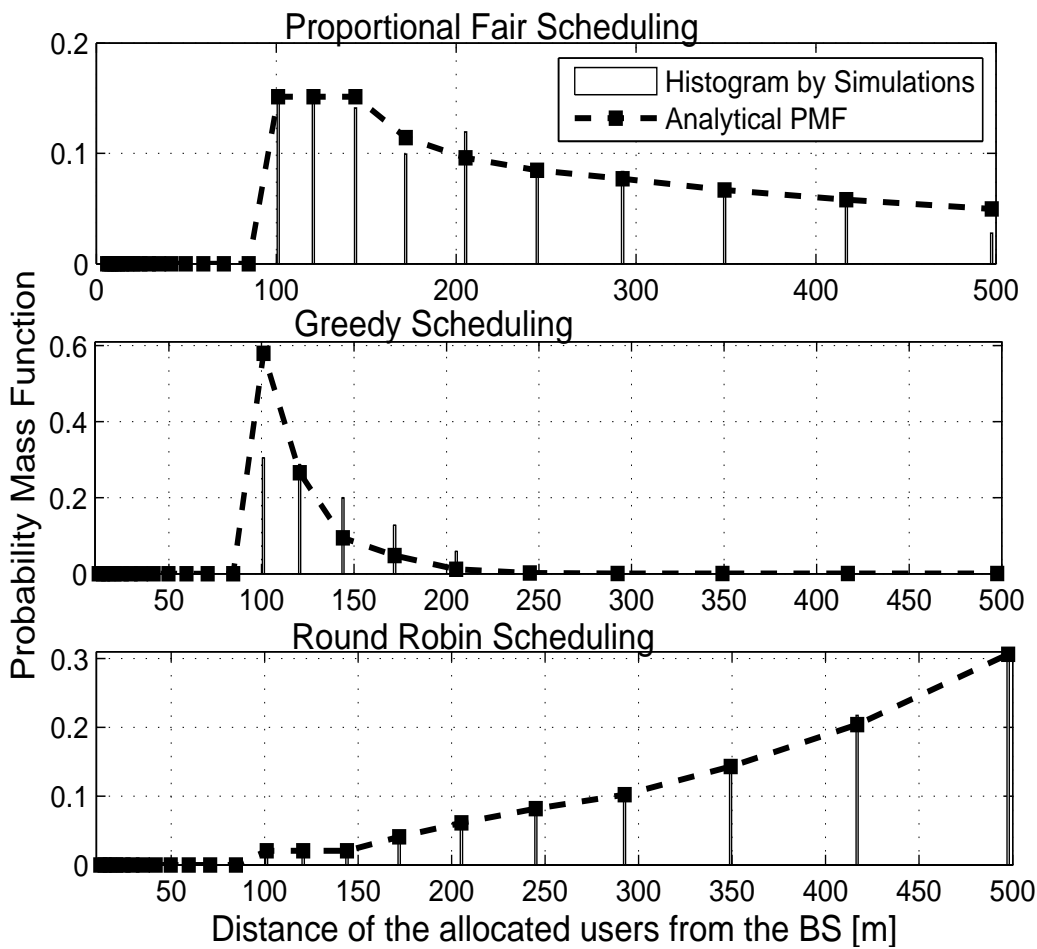}
\caption{PMF of the distance of the allocated users in a given cell (i.e., PMF of $r_{\mathrm{sel}}$) for proportional fair, greedy and round robin scheduling schemes with path loss exponent $\beta=2.6$, $U=50$,  Number of Monte-Carlo simulations =100,000, $C$=60 dB, $P_{\mathrm{max}}$=1W, $\sigma^2$=-174 dBm/Hz.}
\label{Cap1}
\end{figure}

\figref{Cap1} depicts the PMF of the location of the scheduled user in a given cell based on the proportional fair, greedy and round robin scheduling schemes. Since the proportional fair scheme exhibits some fairness among users in a cell, the PMF of the allocated user locations is expected to be more flat compared to the greedy scheme. Since the cell edge has more users due to the large area and each user has equal probability to be allocated on a given subcarrier, therefore the round robin scheme exhibits high probability at the cell-edge. 
It is important to note that the numerical results for the derived PMF in \figref{Cap1} nearly coincide with the exhaustive Monte-Carlo simulation results with a small number of rings $K=10$ and  $\kappa=2$dB. Moreover, it can also be noticed that the width of the circular regions tend to increase from cell center to cell edge which is due to the exponentially decaying path loss as mentioned in Section II. The number of required rings is expected to decrease by reducing $\beta$ and  increasing the amount of power decay within each circular region and vice versa.

Another important point to explain with reference to \figref{Cap1} is that with the increase in the number of competing users on a given subcarrier, the PMF of opportunistic scheduling schemes tends to get skewed which is due to the fact that the higher the  number of users in the cell center, the higher is the probability of allocating a subcarrier in the cell center. In order to get an integer number of users within a ring, we perform rounding in the analysis, i.e., we consider zero active users in the rings where $u_k \leq 0.5$. In Monte-Carlo simulations, we consider the probability of allocating a user in these rings to be zero which can also be verified from \figref{Cap1}. 

\begin{figure}[t]
 \centering
  \includegraphics[totalheight=5in,width=6in]{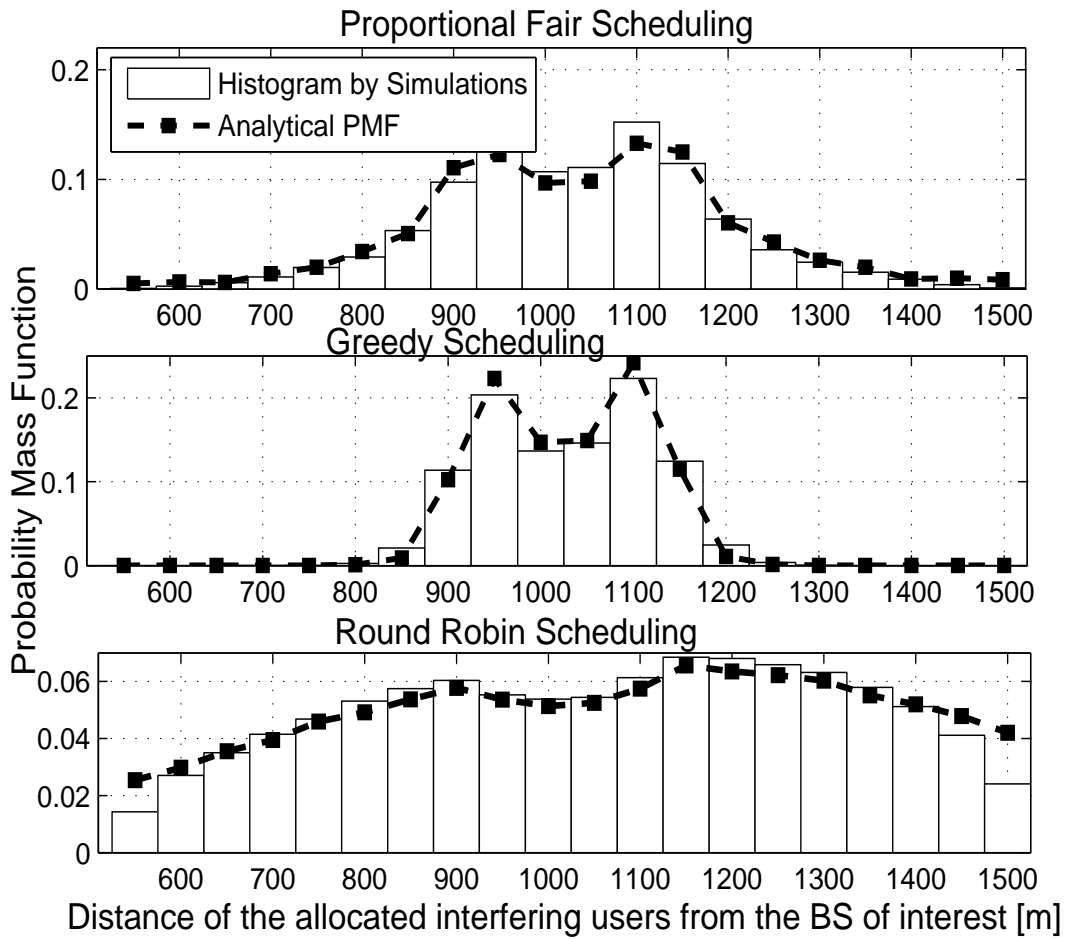}
   \caption{PMF of the distance at which the users in the interfering cells are allocated (i.e., PMF of $\tilde{r}_{\mathrm{sel}}$) for proportional fair, greedy and round robin scheduling schemes with path loss exponent $\beta=2.6$, $U=50$, $\mathcal{I}=720$, $\chi \sim \mathrm{Gamma}(3/2,2/3)$, Number of Monte-Carlo simulations =100,000, $C$=60 dB, $P_{\mathrm{max}}$=1W, $\sigma^2$=-174 dBm/Hz, $\Delta$=50 m.}
\label{Cap2}
\end{figure}

In \figref{Cap2}, the PMF of the distance between the allocated user in interfering cell $l$ and the BS of interest, i.e., $P(\tilde{r}_{\mathrm{sel}}=\tilde{r}_m)$, is presented. Numerical results are found to be in close agreement with the Monte-Carlo simulation results. 
For the opportunistic scheduling schemes, it is likely that a user close to its serving BS can get a subcarrier, thus, the PMF of the distance of allocated interfering users is expected to have high density in the middle. However, the  slight descend in the central region in \figref{Cap2} is due to ignoring users that lie within the rings where the average number of users is less than half.
Moreover, we can observe that the round robin scheduler is highly vulnerable to interference compared to the other schemes as high interference is expected to come from the cell edge users in the interfering cells. On the other hand, the greedy scheduler is expected to have allocations near the cell center and, thus, leads to less interference from neighboring cells. The proportional fair scheme lies in between the two extremes.

\begin{figure}
  \centering
 \includegraphics[totalheight=5in,width=6in]{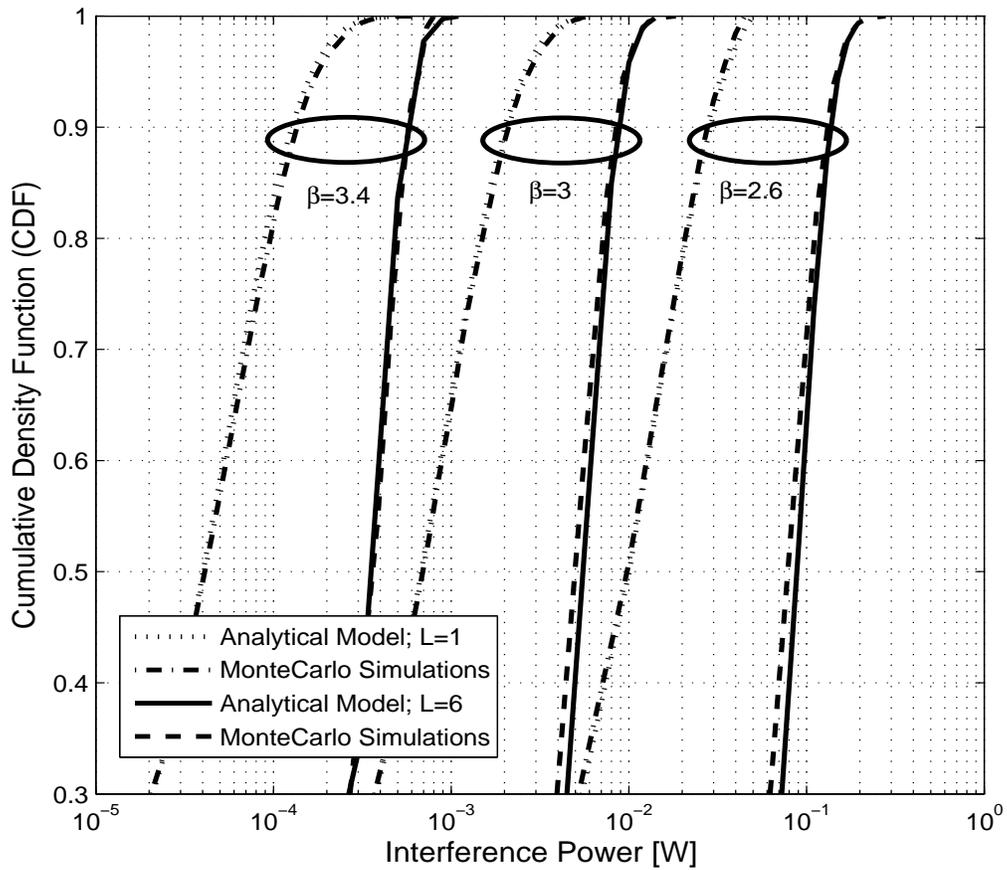}
   \caption{Impact of different number of interferers $L$ and various path loss exponents $(\beta)$ on the CDF of cumulative ICI  considering greedy scheduling scheme, $U=50$, $\mathcal{I}=720$, $\chi \sim \mathrm{Gamma}(3/2,2/3)$, Number of Monte-Carlo simulations =100,000, $C$=60 dB, $P_{\mathrm{max}}$=1W, $\sigma^2$=-174 dBm/Hz.}
\label{Cap3}
\end{figure}

\figref{Cap3} illustrates the CDF of the ICI considering different number of interfering cells and path loss exponents $\beta$ for the greedy scheduling scheme. With the increase in the number of interferers, the interference level increases.  Moreover, as $\beta$ increases, the signal degrades rapidly and thus interference level is reduced considerably.
At this point, it is important to mention  that in this paper we derive and utilize the  MGF of the cumulative ICI rather than the  CDF of the cumulative ICI in order to evaluate important network performance metrics. Therefore, the analytical part of the provided figure of the CDF of the cumulative ICI is plotted   using a  technique mentioned in \cite{sir} to convert MGF into CDF numerically.

\begin{figure}[t]
  \centering
  \includegraphics[totalheight=5in,width=6in]{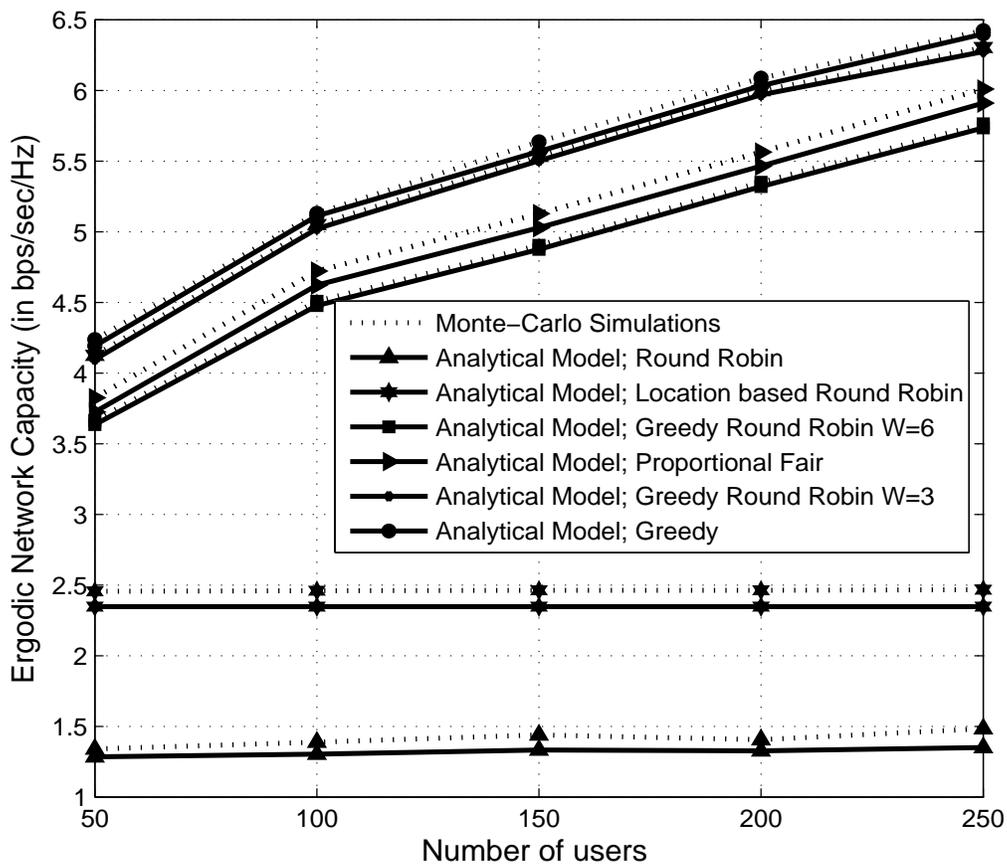}
   \caption{Network ergodic capacity  for different number of users considering different scheduling schemes with $\beta=2.6$, $\chi \sim \mathrm{Gamma}(3/2,2/3)$,$C$=60 dB, $P_{\mathrm{max}}$=1W, $\sigma^2$=-174 dBm/Hz.}
\label{Cap4}
\end{figure}

\figref{Cap4} investigates the effect of increasing the number of competing users on a given sub-carrier considering all scheduling schemes. It can be observed that the increase in the number of users enhances the performance of the opportunistic scheduling schemes due to additional multiuser diversity gains.  
The greedy scheme achieves the best performance whereas the round robin scheme achieves the worst performance. As expected, the proportional fair scheme lies in between the two extremes. The average capacity of location based round robin over $W=K$ time slots has been shown  to be better than the conventional round robin scheme.The average capacity results of the greedy round robin scheme is presented for $W=3$ and $W=6$. Clearly, for $W=1$, the scheme is equivalent to the greedy scheme; however, with the increase of time slots, performance degradation takes place due to the reduction of  multiuser diversity caused by ignoring the users from  previously allocated rings. 

\begin{figure}[t]
  \centering
  \includegraphics[totalheight=5in,width=6in]{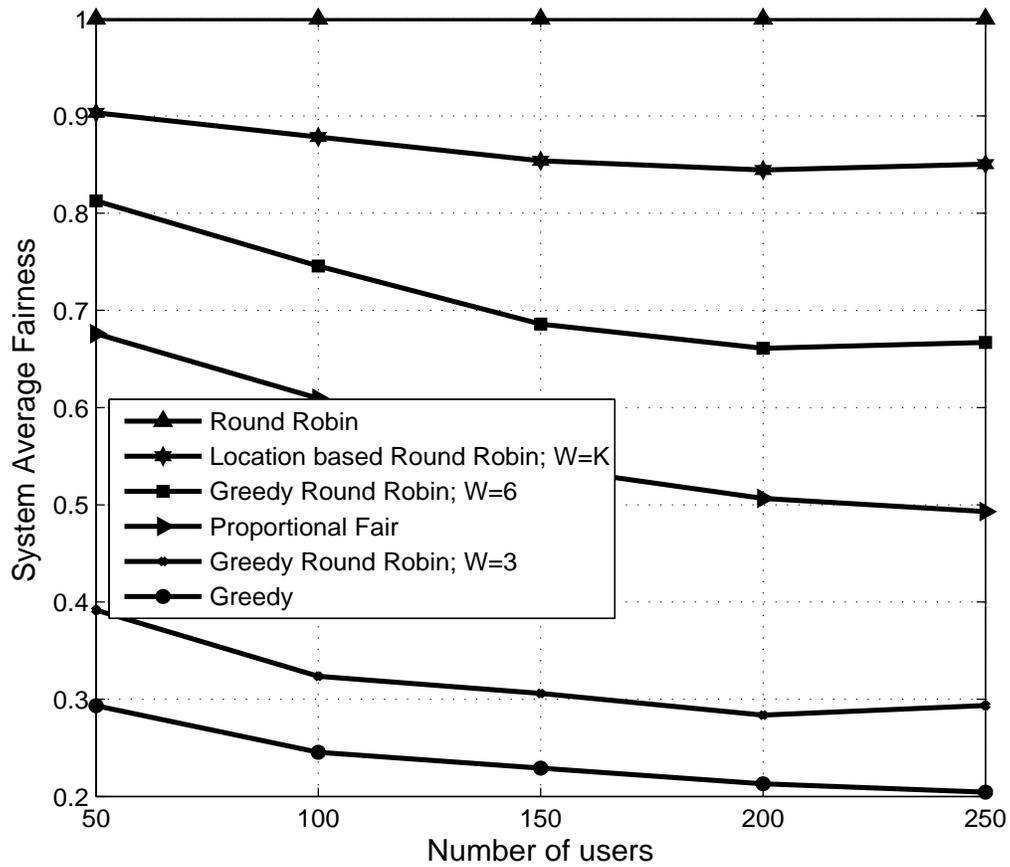}
   \caption{Average system fairness for different number of users considering different scheduling schemes with $\beta=2.6$,$C$=60 dB, $P_{\mathrm{max}}$=1W, $\sigma^2$=-174 dBm/Hz.}
\label{Cap5}
\end{figure}

\figref{Cap5} quantifies the average resource fairness of all  presented scheduling schemes. As expected, round robin is a strictly fair scheme. The proportional fair scheme possesses the ability to enhance the network throughput compared to round robin scheduling while providing a high degree of fairness. The greedy scheme is observed to be the most unfair scheme. Considering $K$ time slots, the average fairness of the location based round robin scheme is investigated and found to be very close to the round robin scheme, however, with degradation in performance as can be observed in \figref{Cap4}. For the greedy round robin scheme, we plotted the fairness metric considering $W=3$ and $W=6$; it is shown that as the number of time slots increases, the fairness improves with a trade-off price in terms of ergodic capacity.

\begin{figure}[t]
  \centering
 \includegraphics[totalheight=5in,width=6in]{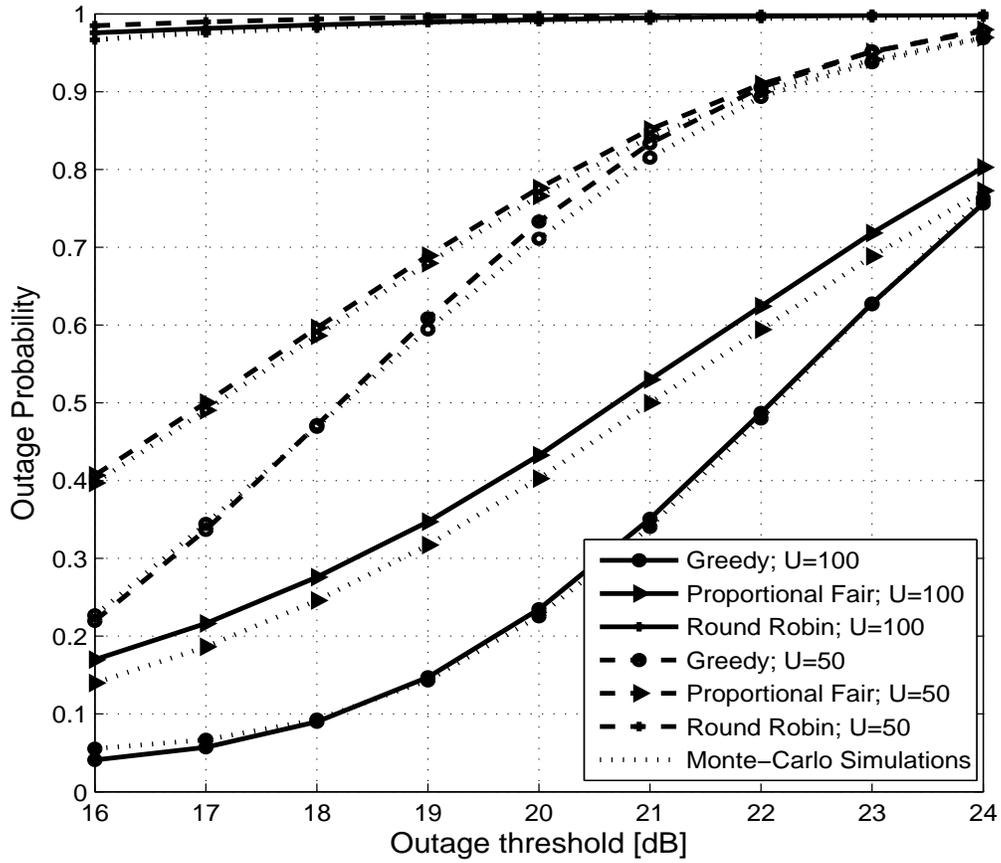}
  \caption{Impact of different scheduling schemes on the network outage probability for different number of users and various  outage thresholds, $\beta$=2.6, $\chi \sim \mathrm{Gamma}(3/2,2/3)$, $C$=60 dB, $P_{\mathrm{max}}$=1W, $\sigma^2$=-174 dBm/Hz.}
\label{Cap6}
\end{figure}

\begin{figure}[t]
  \centering
  \includegraphics[totalheight=5in,width=6in]{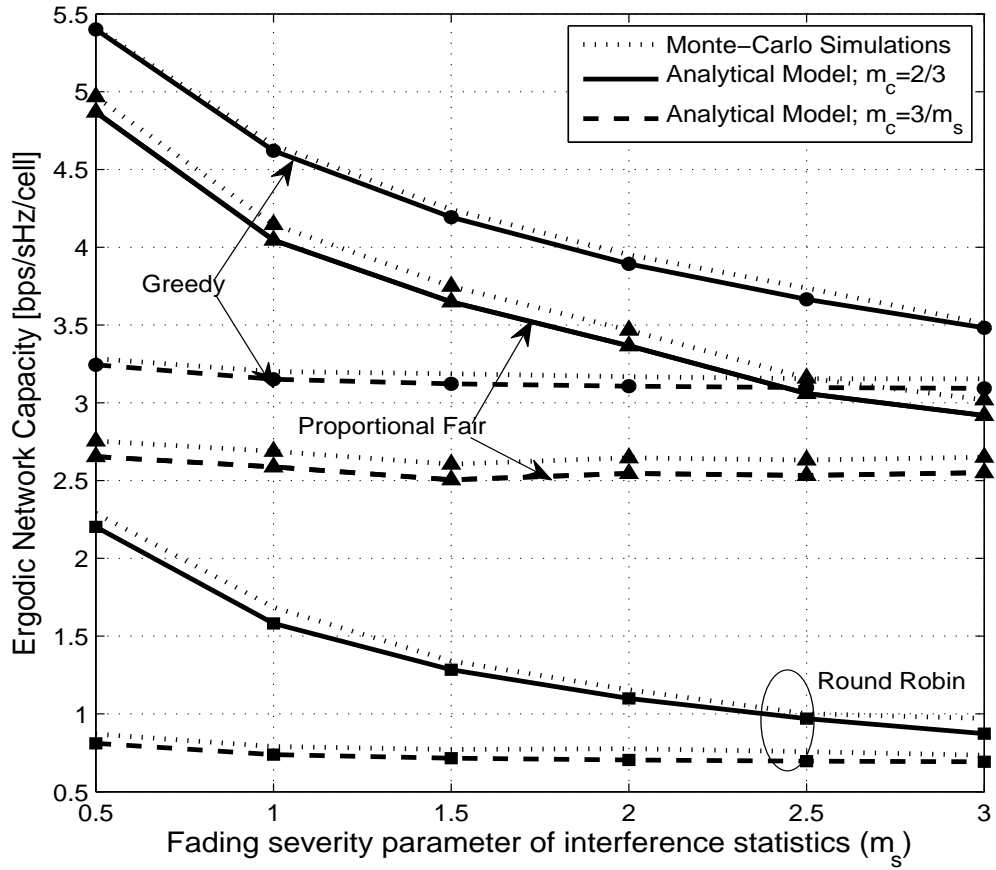}
   \caption{Ergodic network capacity  of different scheduling schemes as a function of  different  parameters of the interference statistics $\chi \sim \mathrm{Gamma} (m_s,m_c)$  with $U =50$ users, $\beta$=2.6, $\sigma^2$=-174 dBm/Hz.}
\label{Cap7}
\end{figure}

In \figref{Cap6}, we evaluate the network outage probability as a function of the outage threshold; $q=(Z+X_0)/Y$ for (i) $U=50$ users; (ii) $U=100$  users. The numerical and simulation results are nearly identical for most cases. The higher the outage threshold for a given  signal and interference power, the greater outage  is expected. Moreover, for larger number of users the outage probability is observed to reduce for all opportunistic scheduling schemes except the round robin scheme. Since increasing the number of users on a given subcarrier in non-opportunistic schemes does not directly affect the access probability of a ring $k$, therefore its impact on the ICI ia almost negligible. This fact can also be verified from \figref{Cap4}.
Finally, in \figref{Cap7}, we evaluate the network ergodic capacity as a function of the fading severity parameter and average power of Gamma fading interference channels for different scheduling schemes. Firstly, it can be observed that  increasing the average power of the interference channel which is given by $\Omega=m_c m_s$ for a given fading severity parameter $m_s$, the capacity degrades significantly for all schemes. Moreover, it is also shown that increasing the fading severity $m_s$ while keeping the average power $\Omega=3$ fixed has minimal impact on the system capacity. Therefore, the lower average power of interference channel $\Omega$, the better is the overall system performance.

\section{Part I: Conclusion}
We proposed a novel approach to model the uplink ICI considering various scheduling schemes and composite fading channel models. The proposed approach is not  dependent on a particular shadowing and fading statistics, hence,  extensions to different models is possible.  The provided numerical results and help in gaining insights into the behavior of ICI considering different scheduling schemes and composite fading models. Moreover, they provide quantitative assessment of the relative performance of various scheduling schemes which is important for  network design and assessment.


\section*{\bf Part II of the Technical Report}

\section*{Part II: Introduction}
Energy efficient wireless communications has been gaining considerable attention  these days mainly  due to  two  major reasons i) dramatically varying global climate \cite{weather1}, and ii) slowly progressing  battery technology \cite{ miao}. In this context, power adaptation has been evolved  as an efficient approach to reduce per capita power consumption, control inter-cell interference (ICI) and increase fairness among users in future generation wireless networks such as Orthogonal Frequency Division Multiple Access (OFDMA).
In OFDMA networks, the system bandwidth is decomposed into orthogonal subcarriers. These subcarriers are adaptively allocated among users within a cell based on a predefined scheduling scheme and user transmit power levels. The allocated users on the same subcarrier in neighboring cells can cause significant uplink ICI depending on their transmit power level and channel conditions with respect to the base station (BS) of interest. 

Most of the recent literature considered the modeling of ICI in the downlink  where the location of interferers is usually deterministic \cite{eurasip,plass}. However, compared to the downlink, the modeling of ICI in the uplink is more challenging due to the arbitrary locations of the interferers and the powers associated with them. Some interesting analytical models for uplink ICI have been presented in \cite{TWC,IEEELetter}; however, none of them considered the impact of channel based scheduling and power adaptation on the uplink ICI.
Recently, in \cite{ISWCS}, we presented a semi-analytical framework to derive the distribution of uplink ICI on a given subcarrier assuming greedy scheduling  without power adaptation.  
In this paper, we generalize the developed semi-analytical framework to incorporate the impact of  power adaptation on the uplink ICI. This power adaptation promises considerable power savings while allowing high degree of fairness among users. Several power adaptation mechanisms are discussed in \cite{powercontrol} such as fast and slow power control, open-loop and closed-loop power control etc. In this paper we focus  on  slow power control  considering that each user is capable of adapting its transmit power autonomously either by measuring its location through a global positioning system (GPS) or estimating its distance based on the power measurement of pilot signals from the surrounding BSs \cite{loc2}.


\section*{Part II: System Model}
We consider a given cell surrounded by $L$ interfering cells. For analytical convenience, the cells are assumed to be circular with radius $R$. Each cell contains $U$  uniformly distributed users where each user is assumed to have perfect knowledge of its distance to the serving BS. The rate adaptation and allocation of users on a given subcarrier, therefore, depend on the channel qualities as well as the transmit powers of the users. The  instantaneous SNR of any user can then be written as:
\begin{equation}
\label{model1}
\gamma = \frac{\mathrm{min} (P_{\mathrm{max}}, P_0 r^\beta) r^{-\beta} \zeta}{\sigma^2}
\end{equation}
where
$P_{\mathrm{max}}$[W] is the maximum transmit power of a user, $P_0$[W] is the desired power level at the receiver, $r$[m] is the user distance from its serving BS, $\beta$ is the path loss exponent, $\sigma^2$ denotes the  thermal noise at the receiver which is considered to be unity  without loss of generality  and $\zeta$  represents the combined shadowing and fading random variable (RV). More explicitly, \eqref{model1} can be re-written as:
\begin{equation}
\label{modelfinal1}
\gamma=
\begin{cases}
P_0 \zeta, & P_0 r^{\beta} < P_{\mathrm{max}}\\
P_{\mathrm{max}} r^{-\beta}\zeta, & P_0 r^{\beta} \geq P_{\mathrm{max}}
\end{cases}
\end{equation}
The distance at which users need their maximum power to compensate path loss completely is referred to as {\it threshold distance }$(r_t)$ and can be computed as follows: 
\begin{equation}
r_t=\left(\frac{P_{\mathrm{max}}}{P_0}\right)^{1/\beta}
\end{equation}
Users located within $r_t$ can compensate path loss completely while saving some proportion of their power, whereas the users located beyond $r_t$  transmit with their maximum power to achieve a certain rate that is less than their desired target.

Each cell is decomposed into $K$ concentric circular rings. The circular regions between two adjacent rings are characterized by uniform path loss variation (in dB) and, thus, possess non-uniform width $\Delta_k$. Since path loss varies exponentially with distance, $\Delta_k$ increases from cell center to cell edge. Thus, the number of circular regions in each cell depends on the path loss exponent. The average number of users in a given ring $k$ can be computed as follows:
\begin{equation}
u_k=\frac{U(r_{k}^2-r_{k-1}^2)}{R^2} \:\:\:\:\: k=1,2,\cdots,K,
\end{equation}
where $r_k$ denotes the radius of ring $k$. It is important to note that $u_k$ can be a fraction of a number; therefore, we round off the fractional part of users in each ring. The motivation behind dividing each cell into a number of circular regions is that in each region the channel conditions of the users become relatively similar especially for large values of $K$.

The proposed approach to model ICI is detailed in the following steps:
\begin{itemize}
\item Derive the distribution $f_{r_{\mathrm{sel}}}(r)$ of allocating a given subcarrier to a user at a distance $r_{\mathrm{sel}}$ from its BS. 
\item Derive the distribution of the distance between the allocated  interfering users and the BS of the cell of interest, i.e., determine $f_{\tilde{r}_{\mathrm{sel}}}(\tilde{r})$ using $f_{r_{\mathrm{sel}}}(r)$ where $\tilde{r}_{\mathrm{sel}}$ is the distance between interfering users and the BS of interest.
\item Derive the distribution of the ICI $f_{X_l}(x)$ from the allocated user in neighboring cell $l$ to the BS of interest. Since the allocated interfering user can transmit with different power levels depending on the distance from its own serving BS,  the incurred interference can be modeled as
\begin{equation}\label{interference}
X_l=
\begin{cases}
P_{0}{r}^{\beta}\tilde{r}^{-\beta} \chi 
&{\Large{\substack {\tilde{r} \in [D-r_t \:\:\: D+r_t], \:{r} \in 
[0\:\:\:   r_t]}}}
\\
P_{\mathrm{max}}\tilde{r}^{-\beta} \chi
& \mathrm{otherwise}
\end{cases}
\end{equation}
where $D=2 R$ and $\chi$ denotes the combined shadowing and fading component of the interference statistics.
\item Derive the MGF of the cumulative interference ${Y}=\sum_{l=1}^{L}X_l$ caused by the allocated interfering users  in all neighboring cells.
\end{itemize}

\section*{Part II: PMF of Allocated User Locations}
In this section, we  derive the discrete distribution of the distance of the allocated users  in a given cell, i.e., the probability mass function (PMF) of $r_{\mathrm{sel}}$. The derivation is divided into two steps explained as follows:
\\
\noindent\textbf{Step~1~(\rm\textit{Selecting the user with the highest SNR in ring $k$}):} 
Since each circular region has uniform path loss variation, the users within a ring $k$ are assumed to be subject to approximately the same path loss. Thus, selecting a user in a ring $k$ is equivalent to selecting the user with maximum channel gain among all the users in ring $k$, i.e., 
\begin{equation}
\zeta_k=\text{max}\{\zeta_1,\zeta_2,\cdots,\zeta_i,\cdots, \zeta_{u_k}\}
\end{equation} 
For simplicity, we consider independent and identically distributed (i.i.d.) channel gains of all users. Therefore, for any ring $k$, the CDF and PDF of the maximum channel gain $\zeta_k$ can be written as follows, respectively:
\begin{equation}
\label{CDF1}
F_{\zeta_k}(\zeta_k)=\prod_{i=1}^{u_k}F_{\zeta_i}(\zeta_k)=\left(F_{\zeta}(\zeta_k)\right)^{u_k}
\end{equation} 
\begin{equation}
\label{PDF1}
f_{\zeta_k}(\zeta_k)=\sum_{j=1}^{u_k} f_{\zeta_j}(\zeta_k)\prod_{i=1,i\neq j}^{u_k}F_{\zeta_i}(\zeta_k)={u_k}f_{\zeta}(\zeta_k)\left(F_{\zeta}(\zeta_k)\right)^{u_k-1}
\end{equation}
Considering the model in  \eqref{modelfinal1}, we split the analysis into two  regions, namely the region within the threshold distance and the region beyond the threshold distance. After performing the RV transformation, we can write the CDF  of the selected user SNR in each ring $k$ ($\gamma_{k}$) as follows:  
\begin{equation}
F_{\gamma_k}(\gamma_k)=
\begin{cases}
\left(F_{\zeta}(\frac{\gamma_k}{P_0})\right)^{u_k}, & r_k < r_t\\
\left(F_{\zeta}(\gamma_{k}r_{k}^{\beta} )\right)^{u_k}, &  r_k \geq r_t
\end{cases}
\end{equation}
\noindent\textbf{Step~2~(\rm\textit{Selecting the ring $k$ with maximum SNR from the $K$ rings}):} 
In this step, we  compute the probability of selecting $k^\mathrm{th}$ ring among all other rings. It is important to note that this is equivalent to selecting the ring $k$ which possesses the user with the highest SNR among all rings. Thus, conditioning on $\gamma_{k}$, the PMF of $r_{\mathrm{sel}}$  can be written as follows:
\begin{equation}
P(r_{\mathrm{sel}}=r_{k}|\gamma_{k})=\prod_{\substack{i=1\\i\neq k}}^{K} F_{\gamma_{i}}(\gamma_{k}) 
\end{equation}
By averaging over  $\gamma_{k}$, the final expression for the PMF of $r_{\mathrm{sel}}$ can be written as follows:
\begin{equation}
\label{IntPMF}
P(r_{\mathrm{sel}}=r_k)=
\int_0^\infty P(r_{\mathrm{sel}}=r_{k}|\gamma_{k})f_{\gamma_{k}}(\gamma_{k}) d\gamma_{k}
\end{equation}
The result in \eqref{IntPMF}  can be evaluated accurately using standard mathematical software packages such as MAPLE and MATHEMATICA and is valid for any composite fading statistics.

Note that $P(r_{\mathrm{sel}}=r_{k})$ in \eqref{IntPMF} is the marginal PMF of $P(r_{\mathrm{sel}}=r_k,\mathcal{\theta}=\theta_n)$ where $\mathcal{\theta}$ is the angular position of the allocated user and is distributed uniformly from $0$ to $2\pi$. Although the RV $\mathcal{\theta}$ possesses continuous distribution, we discretize it in order to reduce complexity. Thus, discretizing $\theta$ into ${N}$ uniform angular intervals, $P(\mathcal{\theta}=\theta_n)$ is ${1}/{N}$, where $\theta_n$ denotes any discrete value that the RV $\mathcal{\theta}$ can take. Since $r_{\mathrm{sel}}$ and $\mathcal{\theta}$ are independent, their joint PMF can be written as:
\begin{equation}
\label{IntPMF1}
P(r_{\mathrm{sel}}=r_k, \mathcal{\theta}=\theta_n)=\frac{P(r_{\mathrm{sel}}=r_k)}{N}
\end{equation}

\section*{Part II: Distribution and MGF of the ICI}
In this section, firstly we find the distribution of the distance of the users allocated in the interfering cell $l$ to the BS of interest, i.e., $f_{\tilde{r}_{sel_l}}(\tilde{r})$. Based on the derived expression we 
derive the distribution of ICI from $l^{\mathrm{th}}$ interfering cell, i.e., $f_{X_l}(x)$ and the MGF of the cumulative ICI $Y$. 
\subsection{Distribution of the allocated interfering user locations}
 Since, each cell is assumed to have identical conditions, $f_{\tilde{r}_{\mathrm{sel}}}(\tilde{r})$ remains the same for all interfering cells and we will not use subscript $l$ any further to simplify notations. Using the cosine law, we can write:
\begin{equation}\label{cosine law}
\begin{split}
\tilde{r}_{{\mathrm{sel}}}^2=r_{{\mathrm{sel}}}^2+D^2-2{r}_{{\mathrm{sel}}}D \;\mathrm{cos}{\theta}
\end{split}
\end{equation}
where
$\tilde{r}_{{\mathrm{sel}}}$ is the distance of the selected interfering user in cell $l$ from the BS of interest, $r_{{\mathrm{sel}}}$ is the distance of the selected interfering user from its own BS, i.e., (BS $l$), $\theta \in \{0, 2\pi\}$. 

In order to determine the PMF of $\tilde{r}_{\mathrm{sel}}$ where $ \tilde{r}_{\mathrm{sel}} \in \{D-R, D+R\} $, first of all we define $\tilde{r}_{n,k}$ for given $\theta_n$ and $r_k$ using \eqref{cosine law} as follows:
\begin{equation}
\begin{split}
\tilde{r}_{n,k}(r,\theta)=\sqrt{r_{k}^2+D^2-2{r_k}D \;\mathrm{cos}{\theta_n}} \:\:\: \forall r_k , \forall \theta_n
\end{split}
\end{equation}

\noindent Clearly $\tilde{r}_{n,k}(r,\theta)$ are the points at which $P(\tilde{r}_{\mathrm{sel}}=\tilde{r}_{n,k})$ can be defined using \eqref{IntPMF} as 
\begin{equation}
P(\tilde{r}_{\mathrm{sel}}=\tilde{r}_{n,k})\\=\frac{P(r_{\mathrm{sel}}=r_k)}{N}
\end{equation}

\subsection{Part II: Distribution of the ICI from one interfering cell}
Since the interfering users can transmit with  different power levels depending on their distance from their serving BS, the interference $X$ can be categorized into two regions mentioned as follows: 
\begin{equation}\label{interference1}
X=
\begin{cases}
P_{0}{r}_{k}^{\beta}\tilde{r}_{n,k}^{-\beta} \chi 
&{\Large{\substack {\tilde{r}_{n,k} \in [D-r_t \:\:\: D+r_t]}}, {r}_{k} \in 
[0\:\:\:   r_t]}
\\
P_{\mathrm{max}}\tilde{r}_{n,k}^{-\beta} \chi
& \mathrm{otherwise}
\end{cases}
\end{equation}
where $\chi$ denotes the interference channel statistics.
The PDF of $X$ conditioned on $\tilde{r}_{n,k}(r,\theta)$ can be determined by RV transformation as follows: 
\begin{equation}
f_{X|\tilde{r}_{n,k}}(x)=
\begin{cases}
\frac{\tilde{r}_{n,k}^{\beta}{f_\chi(\frac{x}{P_0}{r_{k}}^{-\beta}\tilde{r}_{n,k}^{\beta})}}{{P_0 \:{{r}_{k}}^{\beta}}\:} & \tilde{r}_{n,k} \in [D-r_t \:\:\: D+r_t], {r}_{k} \in 
[0\:\:\:   r_t]
\\
\frac{\tilde{r}_{n,k}^{\beta}}{{P_{\mathrm{max}}}}{f_\chi(\frac{x}{P_{\mathrm{max}}}\tilde{r}_{n,k}^{\beta})}
& \mathrm{otherwise}
\end{cases}
\end{equation}
Simply averaging over  $\tilde{r}_{{n,k}}$ and letting $A=\frac{1}{P_0} {\tilde{r}_{n,k}^\beta(r,\theta)} {r}_k^{-\beta}$ and $B=\frac{1}{P_{\mathrm{max}}} {\tilde{r}_{n,k}}^\beta(r,\theta)$ we can write the distribution of interference, i.e., $f_X(x)$ as shown below:
\begin{equation}\label{final}
f_{X}(x)=
\begin{cases}
\sum_{{r}_{k}} \sum_{\theta_n}A\:P(\tilde{r}_{\mathrm{sel}}=\tilde{r}_{n,k}) {f_\chi(Ax)}&
\tilde{r}_{n,k}
\in [D-r_t \:\:\: D+r_t], {r}_{k} \in 
[0\:\:\:   r_t]
\\
\sum_{{r}_{k}} \sum_{\theta_n}
B{P(\tilde{r}_{\mathrm{sel}}=\tilde{r}_{n,k})}{f_\chi(Bx)}
&\mathrm{otherwise}
\end{cases}
\end{equation}
Finally, $f_X(x)$ can be written explicitly as follows:
\begin{equation}\label{final1}
f_{X}(x)=
\sum_{{r}_{k}\in [0 r_t]} \sum_{\theta_n}A\:P(\tilde{r}_{\mathrm{sel}}=\tilde{r}_{n,k}) {f_\chi(Ax)}
+
\sum_{{r}_{k}\notin [0 r_t]} \sum_{\theta_n}
B{P(\tilde{r}_{\mathrm{sel}}=\tilde{r}_{n,k})}{f_\chi(Bx)}
\end{equation}
\subsection*{MGF of the cumulative  ICI}
Computing the distribution of the cumulative ICI $Y$  requires the convolution of the PDF of $L$ RVs $X_l$, $\forall l=1,2,\cdots L$, which is a tedious task for many practical scenarios. To avoid the convolution operations, we utilize an MGF based approach. Since the scheduling scheme is considered to be identical in all cells, the interferers are i.i.d. and therefore the MGF of the cumulative interference $Y$ can  be written as follows:
\begin{equation}\label{main}
\mathcal{M}_{Y}(t)=\prod_{l=1}^{L} \mathcal{M}_{X_l}(t)=\left(\mathcal{M}_{X}(t)\right)^L=\left(\mathbb{E}[e^{tx}]\right)^L
\end{equation}
Looking at the structure of \eqref{final}, we can derive MGF of any composite fading model as
$
\mathcal{M}_{X}(t)=\int_0^\infty e^{tx}f_{X}(x) dx
$
The expression applies to any kind of composite fading models. Due to space limitations we will study only the MGF of  the Gamma composite fading case, i.e., 
we consider a scenario in which shadowing and fading statistics are modeled by a Gamma and Nakagami distribution (also referred as Generalized-K \cite{KG}), respectively. Recently, in  \cite{KG} an accurate approximation of the  Generalized-K RV using moment matching method has been proposed to increase its analytical tractability, i.e., the Generalized-K distribution can be approximated by a simple Gamma distribution \cite{KG}. Therefore in this case $\mathcal{M}_{X}(t))$ can be derived as follows:
\begin{equation}
\label{finalmgfHG}
\mathcal{M}_{X}(t)=
\sum_{\tilde{r}_k
\in [o \:\:\: r_t]} \sum_{\theta}
 \frac{A^{m_s} P(\tilde{r}_{\mathrm{sel}}=\tilde{r}_{n,k})}{\left(A -m_ct\right)^{m_s}}
+
\sum_{\tilde{r}_{k}
\notin [0 \:\:\: r_t]} \sum_{\theta_n} \frac{B^{m_s} P(\tilde{r}_{\mathrm{sel}}=\tilde{r}_{n,k})}{\left(B -m_ct\right)^{m_s}}
\end{equation}
Finally $\mathcal{M}_Y(t)$ can be given simply using \eqref{main}.

\section*{Part II: Evaluation of  Network Performance Metrics}
In this section, we will utilize the derived expressions to evaluate the   network ergodic capacity, average fairness, and average power preservation per user. The evaluation of outage probability is skipped due to space limitations, however, the readers can refer to \cite{ISWCS} for details.
\subsection{ Evaluation of Network Ergodic Capacity}
Using the lemma derived in \cite{hamdi}, the following expression is valid for interference limited systems:
\begin{equation}
\label{lemma1}
\mathbb{E}\left[\mathrm{ln}\left(1+\frac{X_0}{\sum_{l=1}^L{X}_l}\right)\right]=\int_0^\infty \frac{\mathcal{M}_Y(t)-\mathcal{M}_{X_0,Y}(t)}{t} dt
\end{equation}
where, $\mathcal{M}_Y(t)=\mathbb{E}[e^{-t\sum_{l=1}^{L} X_l}]$ is the MGF of the cumulative interference and $\mathcal{M}_{X_0,Y}(t)=\mathbb{E}[e^{-t(X_0+\sum_{l=1}^{L} X_l)}]=\mathbb{E}[e^{-t(X_0+Y)}]$ is the joint MGF of the corresponding signal power of the scheduled user $X_0$ and cumulative interference $Y$. Since $X_0$ and $Y$ are independent, $\mathcal{M}_{X_0,Y}(t)=\mathcal{M}_{X_0}(t)\mathcal{M}_{Y}(t)$. 
The expression  for $\mathcal{M}_{X_0}(t)$  can be given as:
\begin{equation}
\label{HG}
\mathcal{M}_{X_0}(t)=  \int_0^\infty  e^{t x_0} f_{X_0}(x_0) d x_0=\int_0^\infty  t e^{t x_0} F_{X_0}(x_0) d x_0
\end{equation}
where $F_{X_0}(x_0)=\prod_{i=1}^K F_{\gamma_k}(x_0)$, $f_{X_0}(x_0)=\frac{\partial}{\partial x_0}  F_{X_0}(x_0)$. Closed form expressions are also available for $\mathcal{M}_{X_0}(t)$ in the literature \cite{romero}.

\subsection{Evaluation of Average Fairness and Power Preservation}
To measure the degree of fairness among users, we follow the notion developed in \cite{fairness}. The average fairness in a network with $U$ users is defined as
$\mathcal{F}=-\sum_{i=1}^{U} p_i\frac{ \mathrm{log} p_i}{\mathrm{log}U}$
where $p_i$ is the proportion of resources allocated to a user $i$ or the access probability of user $i$. A system is strictly fair if each user has equal probability to access the channel and in such case the average fairness becomes one. The average fairness can be easily computed using our derived results as:
\begin{equation}
\mathcal{F}=-\sum_{k=1}^{K} P(r_{\mathrm{sel}}=r_k) \frac{ \mathrm{log} P(r_{\mathrm{sel}}=r_k)- \mathrm{log} {u_k}}{\mathrm{log}U}
\end{equation}
where $P(r_{\mathrm{sel}}=r_k)$ is given by \eqref{IntPMF}. Moreover, the average  power savings per subcarrier can also be calculated  as follows:
\begin{equation}
\bar{P}=\sum_{r_k=r_1}^{r_t} P(r_{\mathrm{sel}}=r_k) \left(P_{\mathrm{max}}-P_0 r_k^{\beta}\right)
\end{equation}
where $r_t$ denotes the threshold distance.



\section*{Part II: Results and Analysis}
In this section, we aim to validate the accuracy of the derived  expressions through Monte-Carlo simulations. The results are presented for Gamma  composite fading, i.e., $\zeta  \sim \mathrm{Gamma}(1,1)$, $\chi \sim \mathrm{Gamma}(3/2,2/3)$. 
Firstly we will provide a brief overview of the Monte-Carlo simulation setup.
\begin{enumerate}
\item Generate $U$ uniformly distributed users per cell. Each user  has an instantaneous SNR given by \eqref{model1}. Select a user  with maximum instantaneous SNR. Store the distance of the selected user, i.e., $r_{\mathrm{sel}}$ from the serving BS.
\item Compute the distance of the selected user from the BS of interest, i.e., $\tilde{r}_{\mathrm{sel}}$ using cosine law and finally generate the interference using \eqref{interference}.
\item Repeat all steps for a large number of iterations. Generate histogram for the discrete RV $r_{\mathrm{sel}}$ with non-uniform bin widths. 
\end{enumerate}

\begin{figure}[t]
  \centering
  \includegraphics[totalheight=5in,width=6in]{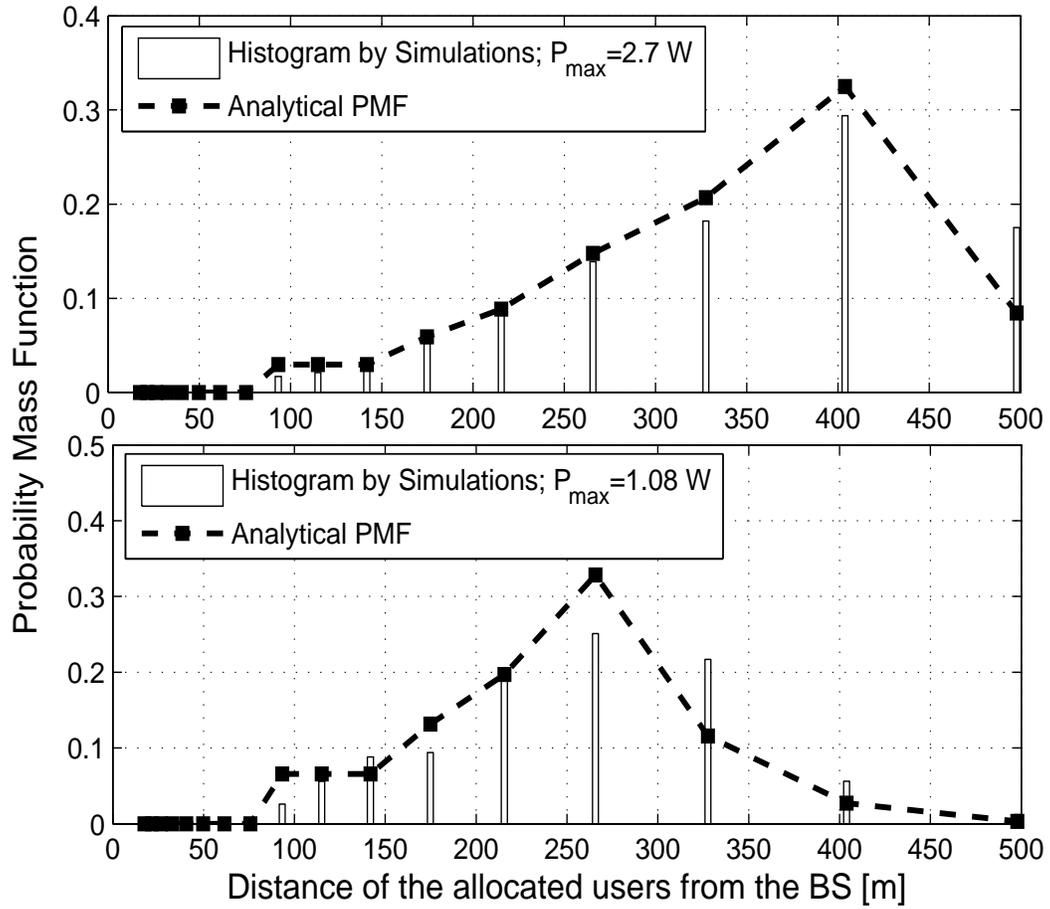}
   \caption{PMF of the distance of allocated users from their serving BS in a given cell (i.e., PMF of $r_{\mathrm{sel}}$) for greedy scheduler with power adaptation, $R$=500 m,  $\beta$=2.2, $U$=50, $P_0$=-23 dBm.}
\label{Cappc0}
\end{figure}

In \figref{Cappc0} the impact of the maximum transmit power limit is shown on the PMF of allocated user locations.  The obtained PMF results fit well with exhaustive Monte-Carlo simulations. Since the users located within the threshold region $r_t$ can  compensate their distance based path loss, each user has on average equal probability of allocation within $r_t$. The increasing trend of PMF within $r_t$  is therefore simply due to an increase  in the number of users in each ring from cell center to the cell edge. It is important to note that the users located beyond $r_t$ transmit with their maximum power as they cannot compensate path loss. These users are therefore scheduled based on their relative channel gains which prioritizes  close users over the far users and hence causes rapid decay of allocation probability beyond  $r_t$. 
In greedy scheduling \cite{ISWCS}, the cell center users have higher priority to be allocated over the cell edge users. On the other side, round robin scheduling provides equal probability of allocation to each user, hence high probability of allocation near the cell edge due to the large area and large number of users at the cell-edge. By observing the result, it can be concluded easily that greedy scheduling with power control (PC) follows the trend of round robin within $r_t$ whereas the trend of greedy scheduling beyond $r_t$. The performance of greedy scheduling with PC is therefore expected to lie in between the two extremes.

Two different transmit power limits are also studied in \figref{Cappc0} which yields two threshold distances, i.e., $r_t=400$m and  $r_t=260 $m, respectively.
It can be observed that the greater the  maximum transmit power, the greater is the threshold distance and more users located farther from the BS become capable to compensate path loss which increases fairness and in turn the incurred ICI.
The slight mismatch in the simulations and analysis  demonstrates the impact of discretization which is dominant for channel based scheduling beyond $r_t$. However, this error  can be reduced by increasing the number of rings.

\begin{figure}[t]
  \centering
  \includegraphics[totalheight=5in,width=6in]{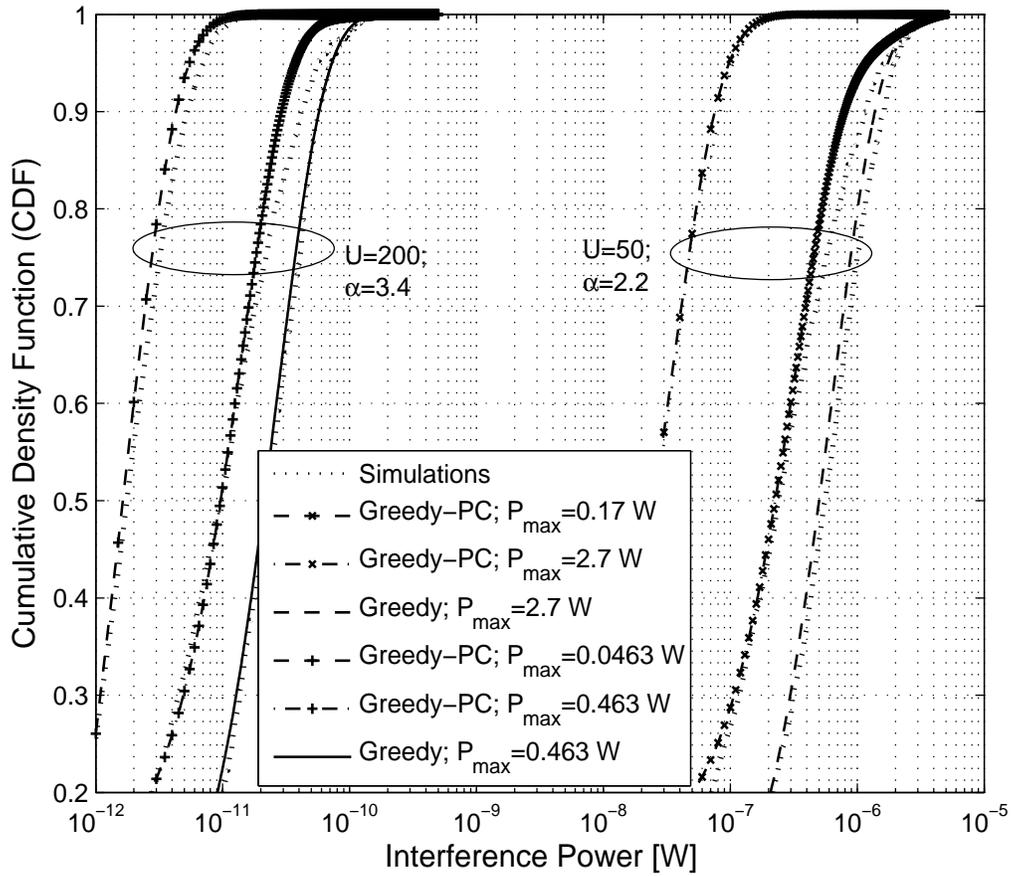}
   \caption{(a) CDF of the ICI for different transmit power levels and path loss exponents considering greedy scheduling with and without power control (PC),  $P_0$ = -23 dBm, $R$ = 500 m.}
\label{Cappc1}
\end{figure}

\begin{figure}[t]
  \centering
  \includegraphics[totalheight=5in,width=6in]{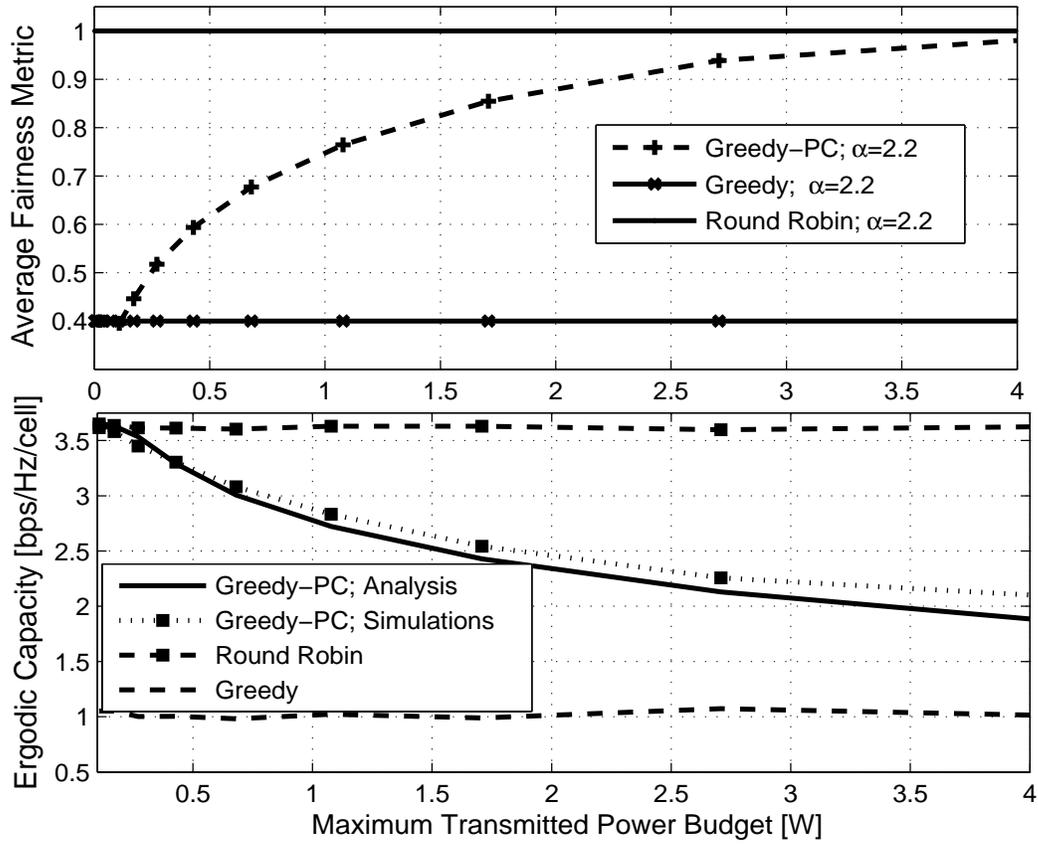}
   \caption{(a) Average fairness among users considering  greedy scheduler with and without power control (PC) and round robin scheduler without PC (b) Ergodic capacity considering  greedy  scheduler with and without PC and round robin scheduler without PC, $U$ = 50, $P_0$ = -23 dBm, $R$ = 500 m, $\beta$ = 2.2.}
\label{Cappc2}
\end{figure}

The CDF of the ICI for different transmit power budgets and different path loss exponents for greedy scheduling with and without PC is plotted in \figref{Cappc1}. High values of path loss exponents causes rapid signal degradation, hence, reduces ICI. Moreover, it can be observed clearly that with low user transmit powers, there is a significant reduction in ICI compared to the high  transmission powers. \emph{It is further interesting to note that the performance of greedy scheduling with PC always remain better than the greedy scheme in terms of incurred ICI, average fairness (see \figref{Cappc2}part(a)) and average power consumption of the users}. 
The top figure in \figref{Cappc2} quantifies the average fairness of the greedy with and without PC and round robin schedulers. 
With the increase of transmit powers, far users  can also adapt their power which increases the average fairness among users. For high user transmit powers, the greedy scheduling with PC achieves the fairness of round robin scheme as is also evident from \figref{Cappc0}. However, it is important to note that the capacity (see \figref{Cappc2}part(b)) and power preservation remains  better than the round robin scheme in which power savings are zero.

The bottom figure in \figref{Cappc2} demonstrate the network capacity of interference limited systems (i.e., thermal noise is neglected). Without PC, the performance of greedy and round robin scheduling remains independent of the transmit power as the factor of $P_{\mathrm{max}}$ cancels out in the capacity calculation.  However, since the greedy scheduling with PC  have less ICI then the greedy scheduler, the network capacity is expected to increase which is not the case as the corresponding user transmit powers are also lowered along with the interfering powers. The main reason of the capacity degradation with the   increase in transmit power budget is that the greater transmission power more users can  compensate  path loss which reduces the number of users transmitting with their maximum powers. This phenomena on one hand increase average  power savings whereas on the other hand degrades overall system capacity.

\bibliography{IEEEfull,references}
\bibliographystyle{IEEEtran}

\newpage
\newpage
\newpage

\newpage
\newpage
\newpage
\newpage

\newpage

\newpage
\newpage

\newpage

\end{document}